\documentclass[11pt,a4paper]{article}
\usepackage{jheppub}

\usepackage{dsfont}
\usepackage{amsmath,amssymb}
\usepackage{graphicx}
\usepackage{axodraw4j}
\usepackage{color}
\usepackage{slashed}

\def\k{{\mathbf k}}
\def\p{{\mathbf p}}
\def\P{{\mathbf P}}
\def\K{{\mathbf K}}
\def\x{{\mathbf x}}
\def\r{{\mathbf r}}

\title{Heavy quarks in proton-nucleus collisions - the hybrid formalism}

\author[a]{Tolga Altinoluk,}
\author[a]{N\'estor Armesto,}
\author[b,c]{Guillaume Beuf,}
\author[d]{Alex Kovner}
\author[b,d]{and Michael Lublinsky}
\affiliation[a]{Departamento de F\'isica de Part\'iculas and IGFAE,
Universidade de Santiago de Compostela,
E-15706 Santiago de Compostela,
Galicia-Spain}
\affiliation[b]{Department of Physics, Ben-Gurion University of the Negev,
Beer Sheva 84105, Israel}
\affiliation[c]{European Centre for Theoretical Studies in Nuclear Physics and Related Areas (ECT*)
and Fondazione Bruno Kessler, Strada delle Tabarelle 286,
 I-38123 Villazzano (TN), Italy}
\affiliation[d]{Physics Department, University of Connecticut, 2152 Hillside
Road, Storrs, CT 06269-3046, USA}

\abstract{We explore the quark mass effects on inclusive hadron  production in proton-nucleus collisions at high energies.
We consider two processes. First, we compute  the single inclusive cross-section for
production of hadrons with open heavy flavour in the proton forward direction { at leading order}.  Next, in the same kinematics, we calculate
{ the heavy-quark contribution to single inclusive production of light or unidentified hadrons at next-to-leading-order}.  For both studies we
exploit the hybrid formalism, that is  the collinear factorisation on the proton side while high-density and high-energy
effects are resummed on the side of the nucleus.
}

\begin{document}
\bibliographystyle{h-physrev4}
\maketitle

%%%%%%%%%%%%%%%%%%%%%%%%%%%%%%%%%%%%%%%%%%%%%%%%%%%%%%%%%%%%%%%%%%%%%%%%%%%%%%%%%%%%%%%%%%%%%%%%%%%%%%%
%%%%%%%%%%%%%%%%%%%%%%%%%%%%%%%%%%%%%%%%%%%%%%%%%%%%%%%%%%%%%%%%%%%%%%%%%%%%%%%%%%%%%%%%%%%%%%%%%%%%%%%
%%%%%%%%%%%%%%%%%%%%%%%%%%%%%%%%%%%%%%%%%%%%%%%%%%%%%%%%%%%%%%%%%%%%%%%%%%%%%%%%%%%%%%%%%%%%%%%%%%%%%%%
%%%%%%%%%%%%%%%%%%%%%%%%%%%%%%%%%%%%%%%%%%%%%%%%%%%%%%%%%%%%%%%%%%%%%%%%%%%%%%%%%%%%%%%%%%%%%%%%%%%%%%%

\section{Introduction and Summary}

Since the original suggestions \cite{glr}, during the last three decades a lot of effort has been devoted to the study of hadronic structure at high energies. The main motivation for it is possible existence of a new regime of Quantum Chromodynamics (QCD) where partonic densities exhibit perturbative saturation. In this regime of  partonic states become dense but the coupling constant is still small and the physics remains perturbative. The recent theoretical implementation of these ideas goes by the name of Color Glass Condensate (CGC) \cite{Mueller,balitsky,balitsky1,Kovchegov,JIMWLK,cgc}. Besides the intense theoretical activity, experiments at the Relativistic Heavy Ion Collider (RHIC) at BNL and the Large Hadron Collider (LHC) at CERN offer new possibilities for searching and characterising such regime.

Admittedly, in spite of the fact that several saturation-based calculations describe data satisfactorily
  (e.g. \cite{KLN,GLLM,cronin,cronin1,LR,DV}), there is no conclusive evidence for the existence of the saturated state in experimental data. One of the main reasons for this is that the accuracy of most calculations is still not sufficient to establish quantitative conclusions. Only a small (although important) part of next-to-leading order (NLO) corrections (the running coupling effects) is presently included in numerical implementations of high-energy evolution \cite{bknumerics} even though the full set of NLO corrections is already available \cite{BKNLO,Grabovsky:2013mba,Balitsky:2013fea,Kovner:2013ona}. Calculation of various observables, like inclusive hadroproduction \cite{AM},
 photoproduction \cite{photo}, etc. is confined  at present to leading order (LO) in the strong coupling constant $\alpha_s$.

Recently several papers have aimed to extend the accuracy of calculations in the CGC framework to NLO: deep inelastic scattering \cite{Balitsky:2010ze,Beuf:2011xd} or single hadroproduction cross section at forward rapidities \cite{production,bowen} in the "hybrid'' formalism \cite{hybrid}. Concerning the latter,
numerical studies indicate very strong effects of the NLO corrections, with cross sections even becoming negative at moderate transverse momenta \cite{amir,ana1}, and even substantial next-to-next-to-leading (NNLO) effects are found \cite{ana2}. More recent discussions focus on the eventual relevance of additional collinear resummations at small $x$  \cite{Beuf:2014uia,Iancu:2015vea} and on the correct choice of the factorisation scale for the high-energy evolution \cite{vitev,Xiao:2014uba}.
In the previous publication \cite{we}, we have introduced a restriction on the lifetime of the partonic fluctuations of the projectile and also revisited the choice of scales. These considerations led to modified NLO expressions which later were shown to improve considerably the stability of the results and their agreement with experimental data \cite{XF2}.

An important aspect of the calculations is  the fact that in hadronic collisions one is forced to rely on factorisation schemes that separate hard scale perturbative processes from the non-perturbative structure
of hadrons. Depending on the kinematics of the process under study, two main factorisation schemes are usually employed. The most common one is  collinear
factorisation \cite{collinear}
(see an introduction to heavy quark production in collinear factorisation in \cite{Martin:2010db}).  In the collinear  scheme, one neglects
the transverse momenta of incoming partons. Production cross sections are computed via a convolution in longitudinal momenta of partonic distribution
functions with hard on-shell  parton production matrix elements. The scheme is normally applicable when produced hadrons have large transverse
momentum, and neglecting the transverse momenta of incoming partons is indeed a valid approximation.
%Relevants Refs \cite{?}....for open charm, $J/\Psi$...
The scheme breaks down, however, when the produced system has relatively low transverse momentum such that either saturation or non-perturbative effects
in  at least  one of the incoming hadrons become important.

An alternative scheme is based on $k_T$ (or, more generally, high-energy) factorisation \cite{ktfact}.
The $k_T$-factorisation scheme is built  upon a separation between hard off-shell matrix elements
and $k_T$-dependent unintegrated gluon densities. It is particularly tuned for central production of states  with relatively small transverse energy.
%{\it There is also a literature of heavy quark production in $k_T$-factorisation \cite{Baranov:2004eu} but with simply off-shell matrix elements from light to heavy, I have never seen a discussion of how to treat heavy quarks as a PDF as they consider semihard production in the central rapidity region at Tevatron and the LHC which should be dominated by the gluons.} H
Heavy quark production within the high-energy factorisation formalism  has been considered e.g. in
Refs. \cite{Baranov:2004eu,GV1,GV2,Tuchin,KovTuch,FGV,Kniehl,FW,KMV, Saleev1,Saleev2,Maciula}.
The scheme breaks down in the forward kinematics, when Bjorken-$x$ of one of the incoming partons is large.

{ Another factorisation scheme, the so called hybrid formalism, was introduced in Ref. \cite{hybrid}}. In the present work we employ the hybrid framework. It is a combination of the previous two factorization approaches
applied to asymmetric production, particularly when the inclusively produced state is measured in the forward direction of one of the colliding hadrons.
In pA collisions, we focus on the hadron production in the proton forward direction. This implies that a relatively large fraction of the proton longitudinal
momentum is taken by the incoming parton, and the collinear factorisation on the proton side can be applied.    At the same time, the target nucleus
is dense. Only its tail of small-$x$  partons, with a typical transverse momenta of order $Q_s$, contributes to particle production.
This is a kinematical regime for which { the  CGC formalism} is most adequate.
Technically, the hybrid formalism is realised in three steps:
\begin{enumerate}
\item A parton is collinearly factorised from the proton. Then,  the order { $g$ contribution to its the light-front wave function
 (including $g \to q\bar q$ splitting in our case) is computed exactly in light-cone perturbation theory \cite{Kogut:1969xa,Bjorken:1970ah,Brodsky:1997de}}.
%, not under the collinear approximation that is employed in DGLAP splitting functions, $k_\perp^2/[z(1-z)(p^{+})^{2}]\ll 1$. As in this case we keep track of the mass that should play an analogous role to $k_\perp$ e.g.
%Guillaume keeps track of the helicity-flip terms in the Dirac algebra, we may be exact in the matrix element. He claims that all he has done is exact in light-front PT.
\item In the CGC, the propagation of gluons  and light quarks is treated eikonally. That is, the $\hat S$-matrix element of all massless partons scattered off the  fast
nucleus is diagonal in  coordinate space and simply given by a light-like Wilson line $U$ in a relevant colour representation. { In the high-energy limit, even massive quarks interact with the target via light-like Wilson lines \cite{Bjorken:1970ah}, up to power-suppressed corrections, due to the Lorentz contraction of the target.
However, when discussing the large mass limit, this approximation can break down. Then, our calculations are valid only as long as the energy of the collision is taken to be much larger than the large mass of the quark.}
% We assume similar light-cone
%propagation for massive quarks too, in spite of their mass.  Assuming Wilson line propagators for heavy quarks is a standard practice employed in even in jet quenching computations of medium-induced radiation off massive quarks  \cite{Armesto:2003jh}.
%We should keep in mind, however that the heavy quark (inifinite mass) limit might not be fully consistent with the eikonal scattering approximation.

\item The partonic-level cross section has to be translated into the hadronic one. This requires a convolution with the { proton} parton distribution and parton fragmentation
functions as in the  usual collinear formalism. On the target nucleus side, we average the Wilson lines with respect to some given distribution $W^T[U]$, as
in all CGC-type calculations.
\end{enumerate}

As discussed above, single inclusive hadron production in the hybrid formalism has been the focus of a large number of recent publications \cite{production,bowen,amir,ana1,ana2,vitev,Xiao:2014uba,we,XF2} . The result
of this series of papers is a   CGC-based computation performed at  a full NLO accuracy { in massless QCD}.  This opens a path for  precise phenomenology based on
saturation physics.
In this paper, we further contribute to this effort by computing the heavy quark contribution { to the NLO correction
for this observable.}
{ We also calculate}
%Particularly, we focus on the calculation at NLO of two channels: (a)
{ single inclusive heavy flavored hadron production at LO,} mainly  $D$ or $B$ mesons (analogous efforts for heavy quarkonia production  can be found in \cite{Fujii,QXF,Xiao3, DLM}).
%; and (b) heavy quark
%contribution to production of light hadrons.
This calculation is relevant for experimental data in
 the forward region. In this respect,
LHCb has measured, in the region $2<y<4.5$ $(5)$, B-meson production in pp collisions at $\sqrt{s}=7$ \cite{Aaij:2012jd,Aaij:2013noa} and 8 \cite{Aaij:2014ija} TeV, $\Lambda_b$ production in pp collisions at $\sqrt{s}=7$ TeV \cite{Aaij:2014jyk}, and prompt charm production at $\sqrt{s}=7$ \cite{Aaij:2013mga} and 13 \cite{Aaij:2015bpa} TeV. ALICE has measured heavy flavour production through its decay into muons \cite{Abelev:2012pi}
in the forward rapidity range $2.5<y<4$.

%We assume no intrinsic charm in the  proton and use  Fixed Flavor Number Scheme.

It is still an open debate how to consistently treat heavy
 flavors in the parton model. The discussion has been mainly conducted in the framework of collinear factorisation and for the initial state.
There are two basic alternatives (see \cite{Martin:2010db,Ball:2015tna,Ball:2015dpa}):
\begin{enumerate}
\item Fixed Flavor Number Scheme ($n$FFNS). $n$ quarks $q$ are considered as massless. Only for massless quarks and gluons there exist parton density functions (PDFs) that evolve according to massless splitting functions, radiation off massive quarks showing no collinear divergence \cite{Dokshitzer:1991fc}. { In most implementations of $n$FFNS, heavy} flavours $Q$ are generated through { gluon splitting  $g \to q\bar q$ and appear at order ${\cal O}(\alpha_s)$.} This scheme should be valid at { moderate scales $\mu$ but collapse when $\mu\gg m_Q$. Indeed, logarithms of $\mu^2/m_Q^2$ that may become large e.g. at large $\mu=p_\perp$, are not resummed.} It is used in some PDF global fits \cite{Alekhin:2013nda}.

\item Variable Flavor Number Scheme (VFNS): $n$ light quarks $q$ are considered as massless. They are evolved as massless up to $\mu^2=m_Q^2$, where the heavy quark PDF appears through a matching, at this scale, to the results of the convolution of the matrix elements to produce heavy flavour with the light flavour PDFs. Above $\mu^2=m_Q^2$, $Q$ is treated as massless for the evolution and there is one additional PDF, so heavy flavour is ${\cal O}(1)$. This scheme { resums} properly the mentioned logarithms, thus it is correct for $\mu^2\gg m_Q^2$, but close to threshold neglects powers of $m_Q^2/\mu^2$. Therefore, a matching of FFNS and VFNS, generically known as Generalized Mass (GM-)VFNS, is nowadays commonly used. There are, at least, three versions of this matching used by different PDF fitting groups in their most recent analysis: TR in MMHT14 \cite{Harland-Lang:2014zoa}, ACOT in CT14 \cite{Dulat:2015mca} and FONLL in NNPDF3.0 \cite{Ball:2014uwa}, see recent discussions in \cite{Ball:2015dpa}. Note that while we have considered the initial state for the discussion, similar considerations hold for fragmentation functions (FFs) \cite{FONLL}. $\alpha_s$ has also to be matched at heavy quark thresholds.

\end{enumerate}

%For our purposes, we will use 3FFNS.
%It  provides charm production and a correction to inclusive light hadron production.
{ As our aim is to provide results valid in the saturation regime for the target, we use the hybrid factorization formalism, and focus on the regime of moderate transverse momentum $\p_h$ of the produced hadron. Hence, logarithms of the type $\ln ({\p_h}^2/m_c^2)$ or $\ln ({\p_h}^2/m_b^2)$ will not be considered large, and accordingly we will use the 3FFNS.}
%As the hydrid formalism is not valid for very large transverse momenta, we expect to be safe of large logarithms $\ln (p_\perp^2/m_c^2)$.
The introduction of  intrinsic charm \cite{Brodsky:1980pb} is a open issue. We will follow the idea in  \cite{Ball:2015dpa} of including it through a non-evolving PDF.
%As long as our $s_0$ is not astronomically larger than $m_c^2$,
%we are kind of safe from the large logs  as the hydrid formalism is valid only for $p_\perp^2\ll s_0$.
%Option 2. looks further more involved, and for GM-VFNS we should look the recipes for matching in the market, or invent ours.

%Intrinsic charm  was only treated, in some versions, by CTEQ is some rough way \cite{Pumplin:2007wg} as kind of an experiment. Note that no one uses massive splitting function (I doubt they exist beyond LO), which besides is kind of senseless as the massive quarks there is no collinear divergence (the dead cone effect \cite{Dokshitzer:1991fc}).

%********

The main results of our paper are the following:
\begin{itemize}
\item We provide results for  the single inclusive cross section for { extrinsic} charm and beauty hadron production { at LO}, given in Eqs. \eqref{hadr_cross_sec_befbef}, \eqref{hadr_cross_sec_aftaft} and \eqref{hadr_cross_sec_interf}.
\item We discuss their heavy quark limit, whose final expression \eqref{hqlimitfinal} is ${\cal O}(1/m_Q^4)$. Interestingly, in the heavy quark limit,
the production cross section is linearly proportional to $Q_s^2$.
\item { We also provide the LO term for the intrinsic contribution to single inclusive heavy flavored hadron production \eqref{hadr_cross_sec_intr_charm_LO}.}
%Our single inclusive cross sections contain the intrinsic heavy flavor contribution, \eqref{hadr_cross_sec_intr_charm_LO}.
\item { We compute the (extrinsic) heavy quark contribution to the NLO corrections to the single inclusive cross section for light or unidentified hadron production, including the heavy quark loop part \eqref{hadr_cross_sec_qq_loop}.}
\end{itemize}

Technical details are provided for all calculations. Note that part of our computations differ from the ones performed in $k_T$ factorisation
  where heavy quarks appear at order $\alpha_s^2$ and, thus, they also contain instantaneous contributions.
We { start with a moderate-$x$ gluon in the proton} and hence our computation is order $\alpha_s$.
{ Individual graphs with a quark loop do have a UV divergence, but these cancel in the sum over the graphs, at the amplitude level.
Moreover, there is no collinear divergence in our calculation, because these are regulated by the heavy quark mass. Nevertheless, we will do the calculation fully in dimensional regularization.
}
%Even though there are no ultraviolet (UV) divergences in our calculations, as they are all regularised by the produced heavy quark mass,
% we carry all the calculations in $D$ dimensions.
Indeed, this facilitates the recovery of the massless limit { in the $\overline{MS}$ factorization scheme, without complicating notably the calculations.
Of course,} it is safe to put $D=4$ in any of our final expressions.
Furthermore, note that since there is no radiation of gluons in the leading-order heavy quark production, no soft divergences appear. Therefore, and at variance with the previous calculation for massless partons \cite{we}, the Ioffe time restriction amounts to a power-suppressed contribution. Since such power suppressed contributions are neglected at the accuracy of our calculation, we do not introduce the Ioffe time restriction in the present paper\footnote{This restriction corresponds to the requirement that the coherence time of the produced fluctuation of the parent parton is larger than the size of the target, in order to guarantee that the fluctuation-target scattering is coherent.}.

%**********

The structure of the paper is as follows. In the next Section (\ref{sec2}), we introduce the light cone perturbation theory applied to a gluon splitting (merging) into a
heavy quark pair. We  compute the amplitude  of the quark pair production in gluon scattering off the nucleus.  Partonic and hadron level cross-sections
are computed in Section \ref{sec3}.  Section  \ref{sec4} is devoted to calculation of heavy quark loop contribution to gluon-to-gluon scattering amplitude. In Section  \ref{sec5},
{ results from previous sections are combined in order to provide the massive quark contribution to the NLO correction to single inclusive light or unidentified} hadron production. Our conventions are summarized in Appendix  \ref{appendix}.

\section{Amplitude for heavy quark-antiquark pair production in gluon scattering on a background field}
\label{sec2}

\subsection{Initial-state gluon wave function including heavy quarks}

\subsubsection{Momentum space}

The Fock state decomposition of the physical (or dressed) state of the incoming gluon, at $x^+=0$, reads (see \cite{Kogut:1969xa} and \cite{Bjorken:1970ah})
%%%%%%%%%%%%%%%%%%%%%%%%%%%%%%%%%%%%%%%%%%%%%%%%%
\begin{eqnarray}
|g(\underline{k_0},\lambda_0,a_0)_{\textrm{phys}}\rangle
&=& \sqrt{Z_A} \bigg[ a^{\dagger}(\underline{k_0},\lambda_0,a_0)\, |0\rangle\nonumber\\
&+&\sum_{q \bar{q}\textrm{ states}}  \Psi_{q_1 \bar{q}_2}^{g_0} \; (t^{a_0})_{\alpha_1\, \alpha_2}\;\; b^{\dagger}(\underline{k_1},h_1,\alpha_1)\, d^{\dagger}(\underline{k_2},h_2,\alpha_2)\, |0\rangle \label{phys_gluon_decomp_mom}\\
&+&\sum_{gg\textrm{ states}}  \Psi_{g_1 g_2}^{g_0} \; (T^{a_0})_{a_1\, a_2} \;\; a^{\dagger}(\underline{k_1},\lambda_1,a_1)\, a^{\dagger}(\underline{k_2},\lambda_2,a_2)\, |0\rangle
+\;\cdots
\bigg].\nonumber
\end{eqnarray}
%%%%%%%%%%%%%%%%%%%%%%%%%%%%%%%%%%%%%%%%%%%%%%%%%
We use the notation $\underline{k}\equiv (k^+, \k)$, and $a^\dagger$, $b^\dagger$ and $d^\dagger$ are creation operators for gluons, quarks and antiquarks respectively. For later convenience, the fundamental ($t^{a_0}$) and adjoint ($T^{a_0}$) color generators have been extracted from the wave functions. Sum over repeated color indices is always implied.
The sums over Fock states contain for each particle the sums over all the quantum numbers (apart from color) and the integration over momentum as
%%%%%%%%%%%%%%%%%%%%%%%%%%%%%%%%%%%%%%%%%%%%%%%%%
\begin{eqnarray}
\int_{0}^{+\infty}\!\!\!\! \frac{dk^+}{(2\pi) 2k^+}
 \int \frac{d^{D-2} \k}{(2\pi)^{D-2}}\ ,
 \label{phase_space_mom}
\end{eqnarray}
%%%%%%%%%%%%%%%%%%%%%%%%%%%%%%%%%%%%%%%%%%%%%%%%%
as well as the symmetry factor $1/n!$ every time that the Fock state contains $n$ identical particles. Hence, there is an $1/2$ factor in the sum over $gg$ states, but not in the one over the $q \bar{q}$ states.

The contribution of gluons and massless quarks to eq. \eqref{phys_gluon_decomp_mom} has already  been calculated, for example in ref. \cite{bowen,we}. Here we are interested in the massive quark contributions. For simplicity, we perform the calculation in the case of QCD with a single massive quark flavor. Indeed, it is trivial to restore the flavor structure at the end of the calculation.

\begin{figure}
\setbox1\hbox to 10cm{
%\fcolorbox{white}{white}{
 \includegraphics{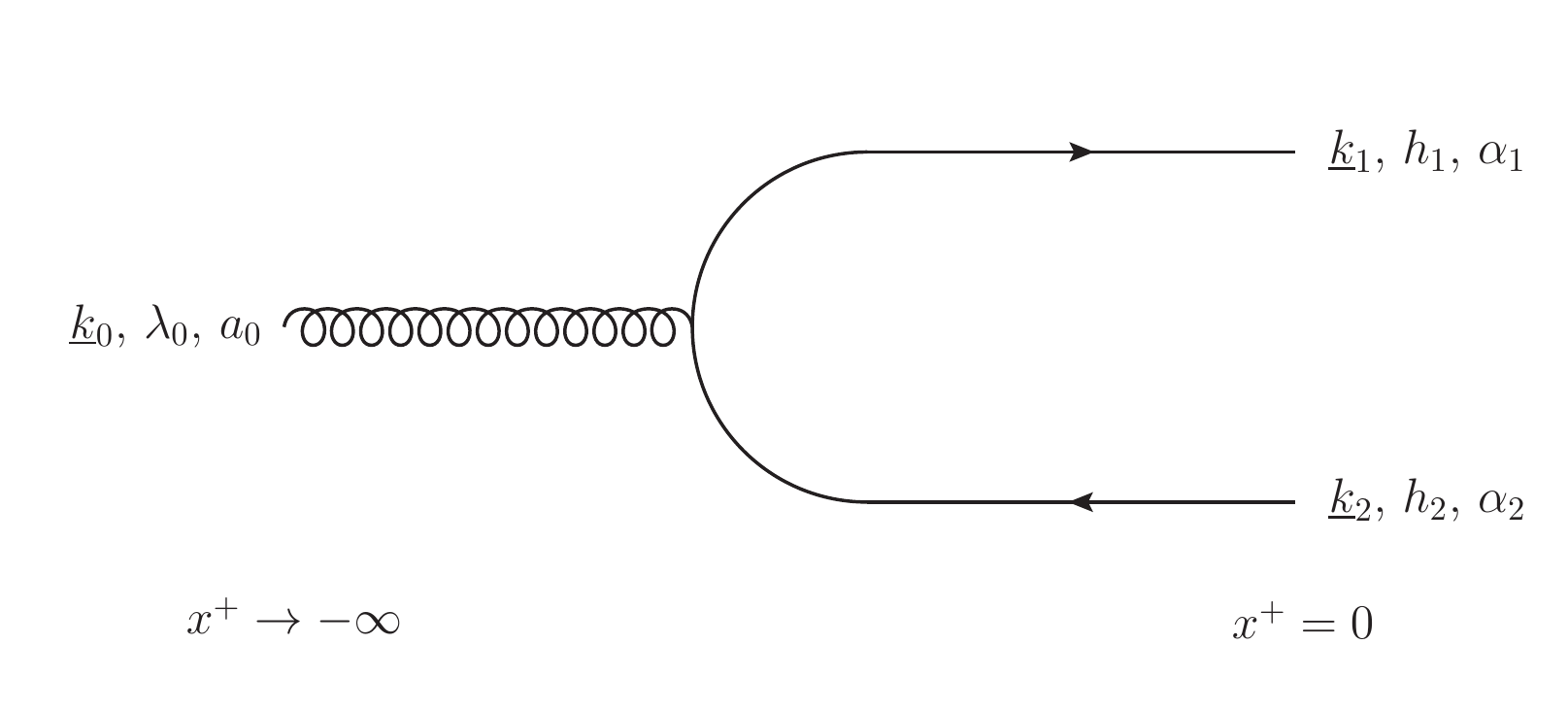}
%}
}
\begin{center}
\hspace{-4cm}
\resizebox*{6cm}{!}{\box1}
\caption{\label{Fig:g2qqbar_IS}Tree-level contribution to the $q\bar{q}$ Fock component of the incoming gluon state. $\lambda_0, a_0$ denote the gluon polarization and color index, and $h_1,h_2$ and $\alpha_1,\alpha_2$ the helicities and color indices of quark and antiquark respectively.}
\end{center}
\end{figure}

At tree level in light-front perturbation theory, there is only one graph, see Fig. \ref{Fig:g2qqbar_IS}, contributing to the $q \bar{q}$ Fock state component of the wave function of the physical  incoming gluon, which gives
%%%%%%%%%%%%%%%%%%%%%%%%%%%%%%%%%%%%%%%%%%%%%%%%%
\begin{eqnarray}
 \Psi_{q_1 \bar{q}_2}^{g_0} \; \; (t^{a_0})_{\alpha_1\, \alpha_2} &=& \frac{\langle0|\, d_2\, b_1\, V_I(0)\, a_0^{\dagger}\, |0\rangle}{\big[k_0^- \!-\! k_1^- \!-\! k_2^- +i \epsilon \big]}\, ,\label{qqbar_inside_g_ISWF}
\end{eqnarray}
%%%%%%%%%%%%%%%%%%%%%%%%%%%%%%%%%%%%%%%%%%%%%%%%%
where $V_I(0)$ is the interaction part of the light-front QCD hamiltonian (see ref. \cite{Brodsky:1997de}) evaluated at $x^+=0$ in the interaction picture.
From the expressions \eqref{quant_free_q} and  \eqref{quant_free_g} of the quantized free fields in the interaction picture, one finds the vertex
%%%%%%%%%%%%%%%%%%%%%%%%%%%%%%%%%%%%%%%%%%%%%%%%%
\begin{eqnarray}
\langle0|\, d_2\, b_1\, V_I(0)\, a_0^{\dagger}\, |0\rangle &=& (2\pi)^{D-1}\delta^{(D-1)}(\underline{k_1}\!+\!\underline{k_2}\!-\!\underline{k_0})\,  (\mu)^{2-\frac{D}{2}}\nonumber\\
&&\times\;  g\, (t^{a_0})_{\alpha_1\, \alpha_2}\; \;
\overline{u}(\underline{k_1},h_1)\; \slashed{\epsilon}_{\lambda_0}\!(\underline{k_0})\; v(\underline{k_2},h_2).
\end{eqnarray}
%%%%%%%%%%%%%%%%%%%%%%%%%%%%%%%%%%%%%%%%%%%%%%%%%

Using the $k^+$ and $\k$ conservation, the energy denominator can be rewritten as
%%%%%%%%%%%%%%%%%%%%%%%%%%%%%%%%%%%%%%%%%%%%%%%%%
\begin{eqnarray}
\big[k_0^- \!-\! k_1^- \!-\! k_2^- +i \epsilon \big]
&=& \bigg[\frac{\k_0^2}{2k_0^+} \!-\! \frac{\k_1^2+m^2}{2k_1^+}\!-\! \frac{\k_2^2+m^2}{2k_2^+} +i \epsilon \bigg]\nonumber\\
&=& -\frac{k_0^+}{2k_1^+\, k_2^+} \bigg[\left(\k_1 \!-\! \frac{k_1^+}{k_0^+}\, \k_0\right)^2 +m^2 -i \epsilon \bigg]
\end{eqnarray}
%%%%%%%%%%%%%%%%%%%%%%%%%%%%%%%%%%%%%%%%%%%%%%%%%
and we can drop the $-i \epsilon $.

Moreover, using relations \eqref{v_bad},  \eqref{ubar_bad} and \eqref{4_to_2_polarization}, one can make explicit all of the transverse momentum dependence of the $q \bar{q} g$ Dirac structure as
%%%%%%%%%%%%%%%%%%%%%%%%%%%%%%%%%%%%%%%%%%%%%%%%%
\begin{eqnarray}
\overline{u}(\underline{k_1},h_1)\; \slashed{\epsilon}_{\lambda_0}\!(\underline{k_0})\; v(\underline{k_2},h_2)
&=&\varepsilon^{j}_{\lambda_0}\; \overline{u_G}(k_1^+,h_1)\bigg[1+ \left(\k_1^i \gamma^i \!+\! m \right)\,  \frac{\gamma^+}{2 k_1^+} \bigg]
\bigg[-\gamma^j+\frac{\k_0^j}{k^+_0}\, \gamma^+\bigg]\nonumber\\
&&\times \, \bigg[1+ \frac{\gamma^+}{2 k_2^+}\, \left(\k_2^l \gamma^l \!-\! m \right) \bigg]
v_G(k_2^+,h_2)\nonumber\\
&&\hspace{-3cm}=-\frac{k_0^+}{2k_1^+\, k_2^+}\, \left(\k_1^i \!-\! \frac{k_1^+}{k_0^+}\, \k_0^i\right) \varepsilon^{j}_{\lambda_0}\; \overline{u_G}(k_1^+,h_1)\, \gamma^+ \bigg[
\frac{(k_0^+ \!-\! 2 k_1^+)}{k_0^+}\, \delta^{ij} + i\, \sigma^{ij}
 \bigg] v_G(k_2^+,h_2)\nonumber\\
&& \hspace{-2.6cm}-\frac{k_0^+}{2k_1^+\, k_2^+}\, m\, \varepsilon^{j}_{\lambda_0}\; \overline{u_G}(k_1^+,h_1)\, \gamma^+ \gamma^j\, v_G(k_2^+,h_2),
\end{eqnarray}
%%%%%%%%%%%%%%%%%%%%%%%%%%%%%%%%%%%%%%%%%%%%%%%%%
where
%%%%%%%%%%%%%%%%%%%%%%%%%%%%%%%%%%%%%%%%%%%%%%%%%
\begin{eqnarray}
\sigma^{ij} &=& \frac{i}{2}\, [\gamma^i,\gamma^j].
\end{eqnarray}
%%%%%%%%%%%%%%%%%%%%%%%%%%%%%%%%%%%%%%%%%%%%%%%%%

So, the tree-level amplitude for $q \bar{q}$ Fock state inside the incoming gluon wave function reads
%%%%%%%%%%%%%%%%%%%%%%%%%%%%%%%%%%%%%%%%%%%%%%%%%
\begin{eqnarray}
 \Psi_{q_1 \bar{q}_2}^{g_0} &=&
 \frac{ (2\pi)^{D-1} \delta^{(D-1)}(\underline{k_1} \!+\!\underline{k_2} \!-\!\underline{k_0})}
 {\bigg[\left(\k_1 \!-\! \frac{k_1^+}{k_0^+}\, \k_0\right)^2 +m^2\bigg]}
 \,  (\mu)^{2-\frac{D}{2}}\, g\, %(t^{a_0})_{\alpha_1\, \alpha_2}
 \nonumber\\
& & \times \Bigg\{ \left(\k_1^i \!-\! \frac{k_1^+}{k_0^+}\, \k_0^i\right) \varepsilon^{j}_{\lambda_0}\; \overline{u_G}(k_1^+,h_1)\, \gamma^+ \bigg[
\frac{(k_0^+ \!-\! 2 k_1^+)}{k_0^+}\, \delta^{ij} + i\, \sigma^{ij}
 \bigg] v_G(k_2^+,h_2)\nonumber\\
& & \hspace{2cm} + m\, \varepsilon^{j}_{\lambda_0}\; \overline{u_G}(k_1^+,h_1)\, \gamma^+ \gamma^j\, v_G(k_2^+,h_2)
 \Bigg\}
 \, .\label{qqbar_inside_g_ISWF_mom}
\end{eqnarray}
%%%%%%%%%%%%%%%%%%%%%%%%%%%%%%%%%%%%%%%%%%%%%%%%%

Compared to the massless case, not only the mass now appears in the denominator, but we also have a new term in the wave function, associated with helicity flip for the quark.

%%%%%%%%%%%%%%%%%%%%%%%%%%%%%%%%%%%%%%%%%%%%%%%%%
%%%%%%%%%%%%%%%%%%%%%%%%%%%%%%%%%%%%%%%%%%%%%%%%%
%%%%%%%%%%%%%%%%%%%%%%%%%%%%%%%%%%%%%%%%%%%%%%%%%
%%%%%%%%%%%%%%%%%%%%%%%%%%%%%%%%%%%%%%%%%%%%%%%%%
%%%%%%%%%%%%%%%%%%%%%%%%%%%%%%%%%%%%%%%%%%%%%%%%%
%%%%%%%%%%%%%%%%%%%%%%%%%%%%%%%%%%%%%%%%%%%%%%%%%

\subsubsection{Mixed space}

Performing the Fourier transform of the Fock states to mixed space, defined by eq. \eqref{FT_a_dag}, we have
%%%%%%%%%%%%%%%%%%%%%%%%%%%%%%%%%%%%%%%%%%%%%%%%%
\begin{eqnarray}
|g_{\textrm{phys}}(\underline{k_0},\lambda_0,a_0)\rangle
&=& \sqrt{Z_A} \bigg[\int d^{D-2}\x_0\;\; e^{i\k_0 \cdot \x_0}\;\;
 a^{\dagger}(k_0^+,\x_0,\lambda_0,a_0)\, |0\rangle\nonumber\\
& & \hspace{-1.7cm}
+\widetilde{\sum_{q \bar{q}\textrm{ states}}}  \widetilde{\Psi}_{q_1 \bar{q}_2}^{g_0}  \;  \; (t^{a_0})_{\alpha_1\, \alpha_2} \;\; \; b^{\dagger}(k_1^+,\x_1,h_1,\alpha_1)\, d^{\dagger}(k_2^+,\x_2,h_2,\alpha_2)\, |0\rangle\label{phys_gluon_decomp_mix}\\
& & \hspace{-1.7cm}
+\widetilde{\sum_{gg\textrm{ states}}}  \widetilde{\Psi}_{g_1 g_2}^{g_0}
 \;  \; (T^{a_0})_{a_1\, a_2}\;\; \; a^{\dagger}(k_1^+,\x_1,\lambda_1,a_1)\, a^{\dagger}(k_2^+,\x_2,\lambda_2,a_2)\, |0\rangle
+\;\cdots
\bigg].\nonumber
\end{eqnarray}
%%%%%%%%%%%%%%%%%%%%%%%%%%%%%%%%%%%%%%%%%%%%%%%%%
The tilde on the sum over Fock states indicates that we replace for each parton the phase space integration eq. \eqref{phase_space_mom} by
%%%%%%%%%%%%%%%%%%%%%%%%%%%%%%%%%%%%%%%%%%%%%%%%%
\begin{eqnarray}
\int_{0}^{+\infty}\!\!\!\! \frac{dk^+}{(2\pi) 2k^+}
 \int d^{D-2} \x\ .
 \label{phase_space_mix}
\end{eqnarray}
%%%%%%%%%%%%%%%%%%%%%%%%%%%%%%%%%%%%%%%%%%%%%%%%%

For the massive quark-antiquark case, the Fourier-transformed amplitude reads
%%%%%%%%%%%%%%%%%%%%%%%%%%%%%%%%%%%%%%%%%%%%%%%%%
\begin{eqnarray}
 \widetilde{\Psi}_{q_1 \bar{q}_2}^{g_0}
 & \equiv & \int \frac{d^{D-2} \k_1}{(2\pi)^{D-2}}
 \int \frac{d^{D-2} \k_2}{(2\pi)^{D-2}} \;
 e^{i\k_1 \cdot \x_1 +i\k_2 \cdot \x_2}\;
 \Psi_{q_1 \bar{q}_2}^{g_0}\nonumber\\
 &=& 2\pi \delta(k_1^+ \!+\!k_2^+ \!-\!k_0^+)\; e^{i\frac{\k_0}{k^+_0}\cdot (k^+_1\, \x_1 +k^+_2 \, \x_2)}\;\;
 (\mu)^{2-\frac{D}{2}}\,  g\, %(t^{a_0})_{\alpha_1\, \alpha_2}
 \nonumber\\
 & & \times \Bigg\{ \varepsilon^{j}_{\lambda_0}\; \overline{u_G}(k_1^+,h_1)\, \gamma^+ \bigg[ \frac{(k_0^+ \!-\! 2 k_1^+)}{k_0^+}\, \delta^{ij} + i\, \sigma^{ij}
 \bigg] v_G(k_2^+,h_2)\;\;
 {\cal B}^i_V(\x_{12},m)\nonumber\\
 & & \hspace{2cm} +m\; \varepsilon^{j}_{\lambda_0}\; \overline{u_G}(k_1^+,h_1)\, \gamma^+ \gamma^j\, v_G(k_2^+,h_2)\;\;
 {\cal B}_S(\x_{12},m)
 \Bigg\}\, , \label{g_to_qqbar_IS_WF}
\end{eqnarray}
%%%%%%%%%%%%%%%%%%%%%%%%%%%%%%%%%%%%%%%%%%%%%%%%%
with the integrals
%%%%%%%%%%%%%%%%%%%%%%%%%%%%%%%%%%%%%%%%%%%%%%%%%
\begin{eqnarray}
{\cal B}^i_V(\x_{12},m) &\equiv & \int \frac{d^{D-2} \K}{(2\pi)^{D-2}}\; e^{i\K \cdot \x_{12}}\; \frac{\K^i}{\K^2+m^2}\nonumber\\
&=& \frac{i}{2\pi}\; \frac{\x_{12}^i}{\x_{12}^2}\, \Big[2\pi\, \x_{12}^2\Big]^{2-\frac{D}{2}}\;\; \Big[m|\x_{12}|\Big]^{\frac{D}{2}-1}\; \mathrm{K}_{\frac{D}{2}-1}\left(m|\x_{12}|\right)\, ,\\
{\cal B}_S(\x_{12},m) &\equiv & \int \frac{d^{D-2} \K}{(2\pi)^{D-2}}\; e^{i\K \cdot \x_{12}}\; \frac{1}{\K^2+m^2}\nonumber\\
&=& \frac{1}{2\pi}\; \Big[2\pi\, \x_{12}^2\Big]^{2-\frac{D}{2}}\;\; \Big[m|\x_{12}|\Big]^{\frac{D}{2}-2}\; \mathrm{K}_{\frac{D}{2}-2}\left(m|\x_{12}|\right)\, .
\end{eqnarray}
%%%%%%%%%%%%%%%%%%%%%%%%%%%%%%%%%%%%%%%%%%%%%%%%%
One recovers the same result as for the $q\bar{q}$ component of the transverse photon wave function (up to the color factor, obviously) as expected, with (in $D=4$) the $\mathrm{K}_{1}$ and $\mathrm{K}_{0}$ modified Bessel functions of the second kind.

%%%%%%%%%%%%%%%%%%%%%%%%%%%%%%%%%%%%%%%%%%%%%%%%%%%%%%%%%%%%%%%%%%%%%%%%%%%%%%%%%%%%%%%%%%%%%%%%%%%%%%%
%%%%%%%%%%%%%%%%%%%%%%%%%%%%%%%%%%%%%%%%%%%%%%%%%%%%%%%%%%%%%%%%%%%%%%%%%%%%%%%%%%%%%%%%%%%%%%%%%%%%%%%
%%%%%%%%%%%%%%%%%%%%%%%%%%%%%%%%%%%%%%%%%%%%%%%%%%%%%%%%%%%%%%%%%%%%%%%%%%%%%%%%%%%%%%%%%%%%%%%%%%%%%%%
%%%%%%%%%%%%%%%%%%%%%%%%%%%%%%%%%%%%%%%%%%%%%%%%%%%%%%%%%%%%%%%%%%%%%%%%%%%%%%%%%%%%%%%%%%%%%%%%%%%%%%%

\subsection{Final state gluon to heavy quark pair splitting}

\subsubsection{Momentum space}

\begin{figure}
\setbox1\hbox to 10cm{
%\fcolorbox{white}{white}{
 \includegraphics{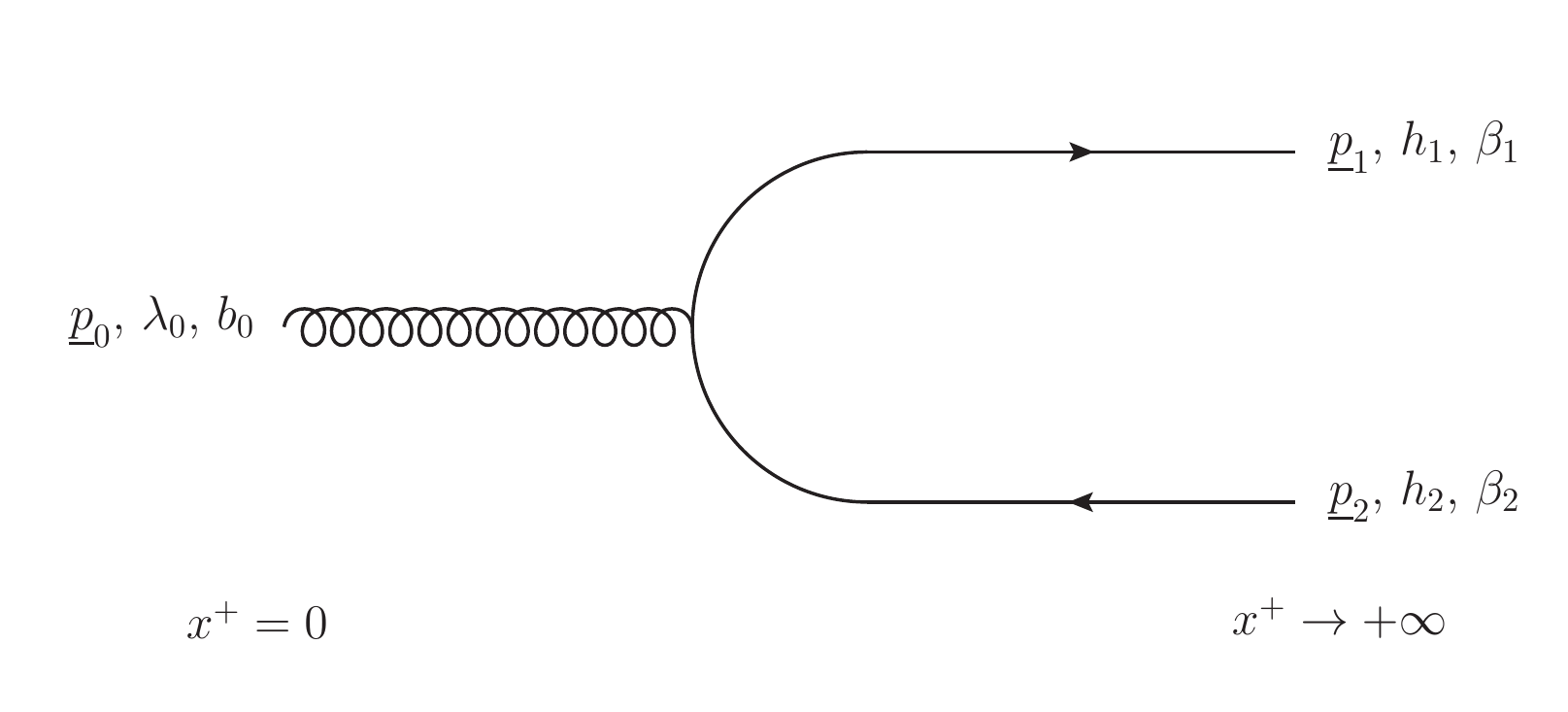}
%}
}
\begin{center}
\hspace{-4cm}
\resizebox*{6cm}{!}{\box1}
\caption{\label{Fig:g2qqbar_FS}Tree-level contribution to the Fock component of the outgoing $q\bar{q}$ state.}
\end{center}
\end{figure}

The Fock state decomposition of the heavy quark-antiquark final state reads
%%%%%%%%%%%%%%%%%%%%%%%%%%%%%%%%%%%%%%%%%%%%%%%%%
\begin{eqnarray}
\langle \bar{q}(\underline{p_2},h_2,\beta_2) q(\underline{p_1},h_1,\beta_1)_{\textrm{phys}}|
&=& \left(\sqrt{Z_{\Psi}}\right)^2 \bigg[\langle 0|\, d(\underline{p_2},h_2,\beta_2)\,
b(\underline{p_1},h_1,\beta_1)\nonumber\\
& & \hspace{-1.2cm}
+\sum_{g\textrm{ states}}  \Phi^{q_1 \bar{q}_2}_{g_0}\;\; (t^{b_0})_{\beta_1\, \beta_2}\;\; \; \langle 0|\, a(\underline{p_0},\lambda_0,b_0)
+\;\cdots
\bigg].
\label{phys_qqbarFS_decomp_mom}
\end{eqnarray}
%%%%%%%%%%%%%%%%%%%%%%%%%%%%%%%%%%%%%%%%%%%%%%%%%
The different terms in this expression have the following interpretation:
\begin{itemize}
  \item First term: trivial contribution with the quark and antiquark directly emerging out of the target at $x^+=0$.
  \item Second term: contribution of gluon splitting to $q\bar{q}$ in the final state, see Fig. \ref{Fig:g2qqbar_FS}.
  \item Other terms: either they are of higher order in $g$, or they will not contribute to the $g+A\rightarrow q+\bar{q}+X$ amplitude in which we are interested.
\end{itemize}

At tree level, only the graph on Fig. \ref{Fig:g2qqbar_FS} contributes to the final state wave function $\Phi^{q_1 \bar{q}_2}_{g_0}$ for the one-gluon Fock component inside the $q\bar{q}$ final state. It gives
%%%%%%%%%%%%%%%%%%%%%%%%%%%%%%%%%%%%%%%%%%%%%%%%%
\begin{eqnarray}
 \Phi^{q_1 \bar{q}_2}_{g_0}\;\; (t^{b_0})_{\beta_1\, \beta_2} &=& \frac{\langle0| d(2)\, b(1)\, V_I(0)\, a^{\dagger}(0)\, |0\rangle }{\big[p_1^- \!+\! p_2^- \!-\! p_0^- +i \epsilon \big]}\, .\label{g_inside_qqbar_FSWF}
\end{eqnarray}
%%%%%%%%%%%%%%%%%%%%%%%%%%%%%%%%%%%%%%%%%%%%%%%%%
Up to the signs in the energy denominator and a trivial relabelling of the momentum variables and color indices, this is identical to its initial state analog $\Psi_{q_1 \bar{q}_2}^{g_0}$, see eq. \eqref{qqbar_inside_g_ISWF}. Hence, from eq. \eqref{qqbar_inside_g_ISWF_mom}
, we deduce
%%%%%%%%%%%%%%%%%%%%%%%%%%%%%%%%%%%%%%%%%%%%%%%%%
\begin{eqnarray}
 \Phi^{q_1 \bar{q}_2}_{g_0}  &=& -
 \frac{ (2\pi)^{D-1} \delta^{(D-1)}(\underline{p_1} \!+\!\underline{p_2} \!-\!\underline{p_0})}
 {\bigg[\left(\p_1 \!-\! \frac{p_1^+}{p_0^+}\, \p_0\right)^2 +m^2\bigg]}
 \,  (\mu)^{2-\frac{D}{2}}\, g\,%(t^{b_0})_{\beta_1\, \beta_2}
 \nonumber\\
& & \times \Bigg\{ \left(\p_1^i \!-\! \frac{p_1^+}{p_0^+}\, \p_0^i\right) \varepsilon^{j}_{\lambda_0}\; \overline{u_G}(p_1^+,h_1)\, \gamma^+ \bigg[
\frac{(p_0^+ \!-\! 2 p_1^+)}{p_0^+}\, \delta^{ij} + i\, \sigma^{ij}
 \bigg] v_G(p_2^+,h_2)\nonumber\\
& & \hspace{2cm} + m\, \varepsilon^{j}_{\lambda_0}\; \overline{u_G}(p_1^+,h_1)\, \gamma^+ \gamma^j\, v_G(p_2^+,h_2)
 \Bigg\}
 \label{g_inside_qqbar_FSWF_mom_1}
\end{eqnarray}
%%%%%%%%%%%%%%%%%%%%%%%%%%%%%%%%%%%%%%%%%%%%%%%%%
and, thus, due to transverse and light-cone momentum conservation,
%%%%%%%%%%%%%%%%%%%%%%%%%%%%%%%%%%%%%%%%%%%%%%%%%
\begin{eqnarray}
 \Phi^{q_1 \bar{q}_2}_{g_0}  &=& -
 \frac{ (2\pi)^{D-1} \delta^{(D-1)}(\underline{p_1} \!+\!\underline{p_2} \!-\!\underline{p_0})}
 {\bigg[\left(\p_2 \!-\! \frac{p_2^+}{p_1^+}\, \p_1\right)^2 +\left(\frac{p_0^+}{p_1^+}\right)^2 m^2\bigg]}
 \,  (\mu)^{2-\frac{D}{2}}\, g\,%(t^{b_0})_{\beta_1\, \beta_2}
 \nonumber\\
& & \times \Bigg\{-\left(\frac{p_0^+}{p_1^+}\right) \left(\p_2^i \!-\! \frac{p_2^+}{p_1^+}\, \p_1^i\right) \varepsilon^{j}_{\lambda_0}\; \overline{u_G}(p_1^+,h_1)\, \gamma^+ \bigg[
\frac{(p_0^+ \!-\! 2 p_1^+)}{p_0^+}\, \delta^{ij} + i\, \sigma^{ij}
 \bigg] v_G(p_2^+,h_2)\nonumber\\
& & \hspace{2cm} + \left(\frac{p_0^+}{p_1^+}\right)^2 m\, \varepsilon^{j}_{\lambda_0}\; \overline{u_G}(p_1^+,h_1)\, \gamma^+ \gamma^j\, v_G(p_2^+,h_2)
 \Bigg\}
 \, .\label{g_inside_qqbar_FSWF_mom_2}
\end{eqnarray}
%%%%%%%%%%%%%%%%%%%%%%%%%%%%%%%%%%%%%%%%%%%%%%%%%

%%%%%%%%%%%%%%%%%%%%%%%%%%%%%%%%%%%%%%%%%%%%%%%%%%%%%%%%%%%%%%%%%%%%%%%%%%%%%%%%%%%%%%%%%%%%%%%%%%%%%%%
%%%%%%%%%%%%%%%%%%%%%%%%%%%%%%%%%%%%%%%%%%%%%%%%%%%%%%%%%%%%%%%%%%%%%%%%%%%%%%%%%%%%%%%%%%%%%%%%%%%%%%%
%%%%%%%%%%%%%%%%%%%%%%%%%%%%%%%%%%%%%%%%%%%%%%%%%%%%%%%%%%%%%%%%%%%%%%%%%%%%%%%%%%%%%%%%%%%%%%%%%%%%%%%
%%%%%%%%%%%%%%%%%%%%%%%%%%%%%%%%%%%%%%%%%%%%%%%%%%%%%%%%%%%%%%%%%%%%%%%%%%%%%%%%%%%%%%%%%%%%%%%%%%%%%%%

\subsubsection{Mixed space}

Rewriting eq. \eqref{phys_qqbarFS_decomp_mom} in mixed space, one gets
%%%%%%%%%%%%%%%%%%%%%%%%%%%%%%%%%%%%%%%%%%%%%%%%%
\begin{eqnarray}
\langle \bar{q}(\underline{p_2},h_2,\beta_2) q(\underline{p_1},h_1,\beta_1)_{\textrm{phys}}|
&=& \left(\sqrt{Z_{\Psi}}\right)^2
\nonumber\\
&&\hspace{-4cm}\times
 \bigg[
\int d^{D-2}\x_1\int d^{D-2}\x_2\;\; e^{-i\p_1 \cdot \x_1-i\p_2 \cdot \x_2}\;
%\nonumber\\
%&&\hspace{1.8cm}\times
\; \langle 0|\, d(p_2^+,\x_2,h_2,\beta_2)\,
b(p_1^+,\x_1,h_1,\beta_1)\nonumber\\
& & \hspace{-1.2cm}
+\widetilde{\sum_{g\textrm{ states}}}  \widetilde{\Phi}^{q_1 \bar{q}_2}_{g_0}\;\;
(t^{b_0})_{\beta_1\, \beta_2}\;\; \; \langle 0|\, a(p_0^+,\x_0,\lambda_0,b_0)
+\;\cdots
\bigg]
\, ,\label{phys_qqbarFS_decomp_mix}
\end{eqnarray}
%%%%%%%%%%%%%%%%%%%%%%%%%%%%%%%%%%%%%%%%%%%%%%%%%
where
%%%%%%%%%%%%%%%%%%%%%%%%%%%%%%%%%%%%%%%%%%%%%%%%%
\begin{eqnarray}
 \widetilde{\Phi}^{q_1 \bar{q}_2}_{g_0}
 & \equiv & \int \frac{d^{D-2} \p_0}{(2\pi)^{D-2}}\;
 e^{-i\p_0 \cdot \x_0}\;
 \Phi^{q_1 \bar{q}_2}_{g_0}\nonumber\\
 &=&
 \frac{- 2\pi \delta(p_1^+ \!+\!p_2^+ \!-\!p_0^+)}
 {\bigg[\left(\p_2 \!-\! \frac{p_2^+}{p_1^+}\, \p_1\right)^2 +\left(\frac{p_0^+}{p_1^+}\right)^2 m^2\bigg]} \; e^{-i(\p_1+\p_2) \cdot \x_0}\;\;
  (\mu)^{2-\frac{D}{2}}\, g\,%(t^{b_0})_{\beta_1\, \beta_2}
 \nonumber\\
 & & \times \Bigg\{-\left(\frac{p_0^+}{p_1^+}\right) \left(\p_2^i \!-\! \frac{p_2^+}{p_1^+}\, \p_1^i\right) \varepsilon^{j}_{\lambda_0}\; \overline{u_G}(p_1^+,h_1)\, \gamma^+ \bigg[
\frac{(p_0^+ \!-\! 2 p_1^+)}{p_0^+}\, \delta^{ij} + i\, \sigma^{ij}
 \bigg] v_G(p_2^+,h_2)\nonumber\\
& & \hspace{2cm} + \left(\frac{p_0^+}{p_1^+}\right)^2 m\, \varepsilon^{j}_{\lambda_0}\; \overline{u_G}(p_1^+,h_1)\, \gamma^+ \gamma^j\, v_G(p_2^+,h_2)
 \Bigg\}\, .
\end{eqnarray}
%%%%%%%%%%%%%%%%%%%%%%%%%%%%%%%%%%%%%%%%%%%%%%%%%

%%%%%%%%%%%%%%%%%%%%%%%%%%%%%%%%%%%%%%%%%%%%%%%%%%%%%%%%%%%%%%%%%%%%%%%%%%%%%%%%%%%%%%%%%%%%%%%%%%%%%%%
%%%%%%%%%%%%%%%%%%%%%%%%%%%%%%%%%%%%%%%%%%%%%%%%%%%%%%%%%%%%%%%%%%%%%%%%%%%%%%%%%%%%%%%%%%%%%%%%%%%%%%%
%%%%%%%%%%%%%%%%%%%%%%%%%%%%%%%%%%%%%%%%%%%%%%%%%%%%%%%%%%%%%%%%%%%%%%%%%%%%%%%%%%%%%%%%%%%%%%%%%%%%%%%
%%%%%%%%%%%%%%%%%%%%%%%%%%%%%%%%%%%%%%%%%%%%%%%%%%%%%%%%%%%%%%%%%%%%%%%%%%%%%%%%%%%%%%%%%%%%%%%%%%%%%%%

%%%%%%%%%%%%%%%%%%%%%%%%%%%%%%%%%%%%%%%%%%%%%%%%%%%%%%%%%%%%%%%%%%%%%%%%%%%%%%%%%%%%%%%%%%%%%%%%%%%%%%%
%%%%%%%%%%%%%%%%%%%%%%%%%%%%%%%%%%%%%%%%%%%%%%%%%%%%%%%%%%%%%%%%%%%%%%%%%%%%%%%%%%%%%%%%%%%%%%%%%%%%%%%
%%%%%%%%%%%%%%%%%%%%%%%%%%%%%%%%%%%%%%%%%%%%%%%%%%%%%%%%%%%%%%%%%%%%%%%%%%%%%%%%%%%%%%%%%%%%%%%%%%%%%%%
%%%%%%%%%%%%%%%%%%%%%%%%%%%%%%%%%%%%%%%%%%%%%%%%%%%%%%%%%%%%%%%%%%%%%%%%%%%%%%%%%%%%%%%%%%%%%%%%%%%%%%%

\subsection{Amplitude for quark-antiquark production in gluon scattering on the background field\label{sec:g2qqbar_ampl}}

We have the mixed-space Fock state decomposition \eqref{phys_gluon_decomp_mix} of the incoming physical gluon. It describes the partonic content of the gluon at $x^+=0$ right before scattering with the target. In the eikonal approximation, the instantaneous interaction with the target is described by the operator $\hat{S}_E$, which introduces a Wilson line for each parton present in the Fock state. More precisely, it acts as
%%%%%%%%%%%%%%%%%%%%%%%%%%%%%%%%%%%%%%%%%%%%%%%%%
\begin{eqnarray}
\hat{S}_E\;  |0\rangle &=& |0\rangle\, , \nonumber\\
\hat{S}_E\;  a^{\dagger}(k^+,\x,\lambda,a) &=& U_A(\x)_{b a}\;\;  a^{\dagger}(k^+,\x,\lambda,b)\; \hat{S}_E\, ,\nonumber\\
\hat{S}_E\;  b^{\dagger}(k^+,\x,h,\alpha) &=& U_F(\x)_{\beta \alpha}\;\;  b^{\dagger}(k^+,\x,h,\beta)\; \hat{S}_E\, ,\nonumber\\
\hat{S}_E\;  d^{\dagger}(k^+,\x,h,\alpha) &=& \Big[U_F^{\dagger}(\x)\Big]_{\alpha \beta}\;\;  d^{\dagger}(k^+,\x,h,\beta)\; \hat{S}_E\, .
\label{action_of_SE}
\end{eqnarray}
%%%%%%%%%%%%%%%%%%%%%%%%%%%%%%%%%%%%%%%%%%%%%%%%%
After applying the operator $\hat{S}_E$ to the initial state \eqref{phys_gluon_decomp_mix}, one only needs to project on the desired final state \eqref{phys_qqbarFS_decomp_mix} in order to get the $S$-matrix element for the massive $q\bar{q}$ production by scattering of a gluon on the target background field. Extracting the delta function ensuring light-cone momentum conservation, we can define the amplitude ${\cal M}_{g\rightarrow q\bar{q}}$ for this process as
%%%%%%%%%%%%%%%%%%%%%%%%%%%%%%%%%%%%%%%%%%%%%%%%%
\begin{eqnarray}
&&\langle \bar{q}(\underline{p_2},h_2,\beta_2) q(\underline{p_1},h_1,\beta_1)_{\textrm{phys}}| \hat{S}_E |g_{\textrm{phys}}(\underline{k_0},\lambda_0,a_0)\rangle \nonumber\\
&&\hskip 2cm = (2k_0^+)(2\pi)\delta(p_1^+\!+\!p_2^+\!-\!k_0^+)\; i\, {\cal M}_{g\rightarrow q\bar{q}}\ .
\end{eqnarray}
%%%%%%%%%%%%%%%%%%%%%%%%%%%%%%%%%%%%%%%%%%%%%%%%%
Then, from the Fock state decompositions \eqref{phys_gluon_decomp_mix} and \eqref{phys_qqbarFS_decomp_mix}, the identities \eqref{action_of_SE} and the commutation relations \eqref{commute_a_adag_mix}, \eqref{anticommute_b_bdag_mix} and \eqref{anticommute_d_ddag_mix}, it is straightforward to calculate the amplitude to leading order in the coupling $g$. One obtains
%%%%%%%%%%%%%%%%%%%%%%%%%%%%%%%%%%%%%%%%%%%%%%%%%
\begin{eqnarray}
i\, {\cal M}_{g\rightarrow q\bar{q}}
&=& i\, {\cal M}^{\textrm{bef}}_{g\rightarrow q\bar{q}}
   +i\, {\cal M}^{\textrm{aft}}_{g\rightarrow q\bar{q}}\ ,
\label{ampl}
\end{eqnarray}
%%%%%%%%%%%%%%%%%%%%%%%%%%%%%%%%%%%%%%%%%%%%%%%%%
where
%%%%%%%%%%%%%%%%%%%%%%%%%%%%%%%%%%%%%%%%%%%%%%%%%
\begin{eqnarray}
i\, {\cal M}^{\textrm{bef}}_{g\rightarrow q\bar{q}}
&=& \frac{1}{2k_0^+}\;  (\mu)^{2-\frac{D}{2}}\, g\,
\int d^{D-2} \x_1\; e^{-i\x_1\cdot \left[\p_1-\frac{p_1^+}{k^+_0}\, \k_0\right]}\;
\int d^{D-2} \x_2\; e^{-i\x_2\cdot \left[\p_2-\frac{p_2^+}{k^+_0}\, \k_0\right]}\nonumber\\
&&\hspace{-2cm} \times \Big[U_F(\x_1)\, t^{a_0}\, U_F^{\dagger}(\x_2)\Big]_{\beta_1\, \beta_2}
%\nonumber\\
\!\varepsilon^{j}_{\lambda_0} \Bigg\{{\cal B}^i_V(\x_{12},m)\; \; \overline{u_G}(p_1^+,h_1)\, \gamma^+ \bigg[ \frac{(k_0^+ \!-\! 2 p_1^+)}{k_0^+}\, \delta^{ij} + i\, \sigma^{ij}
 \bigg] v_G(p_2^+,h_2)
 \nonumber\\
 & & \hspace{3.1cm} +{\cal B}_S(\x_{12},m)\;\; m\; \overline{u_G}(p_1^+,h_1)\, \gamma^+ \gamma^j\, v_G(p_2^+,h_2)
 \Bigg\}
  \label{bef_ampl}
\end{eqnarray}
%%%%%%%%%%%%%%%%%%%%%%%%%%%%%%%%%%%%%%%%%%%%%%%%%
and
%%%%%%%%%%%%%%%%%%%%%%%%%%%%%%%%%%%%%%%%%%%%%%%%%
\begin{eqnarray}
i\, {\cal M}^{\textrm{aft}}_{g\rightarrow q\bar{q}}
&=&- \frac{1}{2k_0^+}\;  (\mu)^{2-\frac{D}{2}}\, g\, \;
\int d^{D-2} \x_0\;  e^{-i\x_0 \cdot (\p_1+\p_2-\k_0)}\;\;
\frac{ (t^{b_0})_{\beta_1\, \beta_2}\; U_A(\x_0)_{b_0\, a_0}}
 {\bigg[\left(\p_2 \!-\! \frac{p_2^+}{p_1^+}\, \p_1\right)^2 +\left(\frac{k_0^+}{p_1^+}\right)^2 m^2\bigg]}\nonumber\\
 & & \times \varepsilon^{j}_{\lambda_0}\;  \Bigg\{-\left(\frac{k_0^+}{p_1^+}\right) \left(\p_2^i \!-\! \frac{p_2^+}{p_1^+}\, \p_1^i\right) \overline{u_G}(p_1^+,h_1)\, \gamma^+ \bigg[
\frac{(k_0^+ \!-\! 2 p_1^+)}{k_0^+}\, \delta^{ij} + i\, \sigma^{ij}
 \bigg] v_G(p_2^+,h_2)\nonumber\\
& & \hspace{2cm} + \left(\frac{k_0^+}{p_1^+}\right)^2 m\; \overline{u_G}(p_1^+,h_1)\, \gamma^+ \gamma^j\, v_G(p_2^+,h_2)
 \Bigg\}
 \label{aft_ampl}
\end{eqnarray}
%%%%%%%%%%%%%%%%%%%%%%%%%%%%%%%%%%%%%%%%%%%%%%%%%
correspond to the contributions where the gluon splits into $q\bar{q}$ before or after crossing the target, respectively.

%%%%%%%%%%%%%%%%%%%%%%%%%%%%%%%%%%%%%%%%%%%%%%%%%%%%%%%%%%%%%%%%%%%%%%%%%%%%%%%%%%%%%%%%%%%%%%%%%%%%%%%
%%%%%%%%%%%%%%%%%%%%%%%%%%%%%%%%%%%%%%%%%%%%%%%%%%%%%%%%%%%%%%%%%%%%%%%%%%%%%%%%%%%%%%%%%%%%%%%%%%%%%%%
%%%%%%%%%%%%%%%%%%%%%%%%%%%%%%%%%%%%%%%%%%%%%%%%%%%%%%%%%%%%%%%%%%%%%%%%%%%%%%%%%%%%%%%%%%%%%%%%%%%%%%%
%%%%%%%%%%%%%%%%%%%%%%%%%%%%%%%%%%%%%%%%%%%%%%%%%%%%%%%%%%%%%%%%%%%%%%%%%%%%%%%%%%%%%%%%%%%%%%%%%%%%%%%

\section{Heavy flavored hadron production in the hybrid factorization}
\label{sec3}

%%%%%%%%%%%%%%%%%%%%%%%%%%%%%%%%%%%%%%%%%%%%%%%%%%%%%%%%%%%%%%%%%%%%%%%%%%%%%%%%%%%%%%%%%%%%%%%%%%%%%%%
%%%%%%%%%%%%%%%%%%%%%%%%%%%%%%%%%%%%%%%%%%%%%%%%%%%%%%%%%%%%%%%%%%%%%%%%%%%%%%%%%%%%%%%%%%%%%%%%%%%%%%%

\subsection{Partonic cross section for heavy quark production}

From the amplitude ${\cal M}_{g\rightarrow q\bar{q}}$, one can obtain the partonic cross section for the process $g+A\rightarrow q+\bar{q}+X$ at LO, as (see ref. \cite{Bjorken:1970ah})
%%%%%%%%%%%%%%%%%%%%%%%%%%%%%%%%%%%%%%%%%%%%%%%%%
\begin{eqnarray}
\hspace{-1cm}(2p_1^+) (2p_2^+)(2\pi)^{2D\!-\!2} \frac{d\sigma^{g+A\rightarrow q+\bar{q}+X}}{dp_1^+\, d^{D\!-\!2}\p_1\, dp_2^+\, d^{D\!-\!2}\p_2}
&=&  (2k^+_0)(2\pi)\delta(p_1^+\!+\!p_2^+\!-\! k_0^+)\;
\frac{1}{d_A}\nonumber\\
&&\times\,
\sum_{a_0,\, \beta_1,\, \beta_2} \frac{1}{D\!-\!2}\sum_{\lambda_0,\, h_1,\, h_2}  \Big|{\cal M}_{g\rightarrow q\bar{q}}\Big|^2
\, ,
\end{eqnarray}
%%%%%%%%%%%%%%%%%%%%%%%%%%%%%%%%%%%%%%%%%%%%%%%%%
where, as usual, one has to sum over the colors and polarizations of final particles and average over the  colors and polarizations of initial particles (note that $D-2$ is indeed the number of physical gluon polarizations in conventional dimensional regularization). $d_A$ is the dimension of the adjoint representation of the gauge group i.e. $d_A=N_c^2\!-\!1$ for $SU(N_c)$.

Then, the single inclusive massive quark production cross section is obtained by integrating over the kinematics of the antiquark, as
%%%%%%%%%%%%%%%%%%%%%%%%%%%%%%%%%%%%%%%%%%%%%%%%%
\begin{eqnarray}
(2p_1^+)(2\pi)^{D\!-\!1} \frac{d\sigma^{g+A\rightarrow q+X}}{dp_1^+\, d^{D\!-\!2}\p_1}
&=&  \int_{0}^{+\infty}\!\!\!\! \frac{dp_2^+}{(2\pi) 2p_2^+}
 \int \frac{d^{D-2} \p_2}{(2\pi)^{D-2}}\;\;
  (2p_1^+) (2p_2^+)(2\pi)^{2D\!-\!2} \nonumber\\
  &&\times \, \frac{d\sigma^{g+A\rightarrow q+\bar{q}+X}}{dp_1^+\, d^{D\!-\!2}\p_1\, dp_2^+\, d^{D\!-\!2}\p_2}\, .
\label{pair_to_single_incl}
\end{eqnarray}
%%%%%%%%%%%%%%%%%%%%%%%%%%%%%%%%%%%%%%%%%%%%%%%%%

\subsubsection{Spin}

Note that both contributions to the amplitude, \eqref{bef_ampl} and \eqref{aft_ampl} are linear combinations of the same two spinor structures (spin-flip and spin non-flip), which contain all of the dependence on the helicities of the quarks, whereas the dependence on the gluon polarization always appears via a $\varepsilon^{j}_{\lambda_0}$ factor.
The spin sum/average for the square of the spin-flip structure reads
%%%%%%%%%%%%%%%%%%%%%%%%%%%%%%%%%%%%%%%%%%%%%%%%%
\begin{eqnarray}
&& \frac{1}{D\!-\!2}\sum_{\lambda_0,\, h_1,\, h_2} \varepsilon^{j'\, *}_{\lambda_0}
  \varepsilon^{j}_{\lambda_0}
  \bigg[\overline{u_G}(p_1^+,h_1)\, \gamma^+ \gamma^{j'}\, v_G(p_2^+,h_2)\bigg]^{\dagger}
  \overline{u_G}(p_1^+,h_1)\, \gamma^+ \gamma^j\, v_G(p_2^+,h_2)\nonumber\\
&=&   \frac{-g^{jj'}}{D\!-\!2}\; \sum_{h_1,\, h_2}
- \: \overline{v_G}(p_2^+,h_2)\, \gamma^+ \gamma^{j'}\, u_G(p_1^+,h_1)
 \overline{u_G}(p_1^+,h_1)\, \gamma^+ \gamma^j\, v_G(p_2^+,h_2)\nonumber\\
&=&   \frac{-g^{jj'}}{D\!-\!2}\; (-1)(2p_1^+)(2p_2^+)\;\;  \textrm{Tr}\big[{{\cal P}_{G}}\, \gamma^{j'}\, \gamma^{j} \big]\nonumber\\
&=&   2(2p_1^+)(2p_2^+)\, .
\end{eqnarray}
%%%%%%%%%%%%%%%%%%%%%%%%%%%%%%%%%%%%%%%%%%%%%%%%%
For the square of the spin non-flip structure, one gets
%%%%%%%%%%%%%%%%%%%%%%%%%%%%%%%%%%%%%%%%%%%%%%%%%
\begin{eqnarray}
&& \frac{1}{D\!-\!2}\sum_{\lambda_0,\, h_1,\, h_2} \varepsilon^{j'\, *}_{\lambda_0}
  \varepsilon^{j}_{\lambda_0}
  \bigg[\overline{u_G}(p_1^+,h_1)\, \gamma^+ \bigg[
\frac{(k_0^+ \!-\! 2 p_1^+)}{k_0^+}\, \delta^{i'j'} + i\, \sigma^{i'j'}
 \bigg] v_G(p_2^+,h_2)\bigg]^{\dagger}\nonumber\\
 && \hspace{1cm}\times \,
  \overline{u_G}(p_1^+,h_1)\, \gamma^+ \bigg[
\frac{(k_0^+ \!-\! 2 p_1^+)}{k_0^+}\, \delta^{ij} + i\, \sigma^{ij}
 \bigg]\, v_G(p_2^+,h_2)\nonumber\\
&=&   \frac{-g^{jj'}}{D\!-\!2}\; (2p_1^+)(2p_2^+)
\textrm{Tr}\bigg[{{\cal P}_{G}}\, \bigg(
\frac{(k_0^+ \!-\! 2 p_1^+)}{k_0^+}\, \delta^{i'j'} - i\, \sigma^{i'j'}
 \bigg)\, \bigg(
\frac{(k_0^+ \!-\! 2 p_1^+)}{k_0^+}\, \delta^{ij} + i\, \sigma^{ij}
 \bigg) \bigg]
 \nonumber\\
&=&   4(2p_1^+)(2p_2^+)\frac{\big(-g^{ii'}\big)}{D\!-\!2}\; \bigg[\left(\frac{p_1^+}{k_0^+}\right)^2+\left(\frac{p_2^+}{k_0^+}\right)^2 +\frac{D\!-\!4}{2}
\bigg]
\, .
\end{eqnarray}
%%%%%%%%%%%%%%%%%%%%%%%%%%%%%%%%%%%%%%%%%%%%%%%%%
For the interference between the spin flip and spin non-flip structures, the helicity sums lead to the trace of an odd number of gamma matrices, and thus vanish. Hence, as expected, there is no interference between the spin flip and spin non-flip contributions to the amplitude ${\cal M}_{g\rightarrow q\bar{q}}$.

%%%%%%%%%%%%%%%%%%%%%%%%%%%%%%%%%%%%%%%%%%%%%%%%%%%%%%%%%%%%%%%%%%%%%%%%%%%%%%%%%%%%%%%%%%%%%%%%%%%%%%%
%%%%%%%%%%%%%%%%%%%%%%%%%%%%%%%%%%%%%%%%%%%%%%%%%%%%%%%%%%%%%%%%%%%%%%%%%%%%%%%%%%%%%%%%%%%%%%%%%%%%%%%
%%%%%%%%%%%%%%%%%%%%%%%%%%%%%%%%%%%%%%%%%%%%%%%%%%%%%%%%%%%%%%%%%%%%%%%%%%%%%%%%%%%%%%%%%%%%%%%%%%%%%%%
%%%%%%%%%%%%%%%%%%%%%%%%%%%%%%%%%%%%%%%%%%%%%%%%%%%%%%%%%%%%%%%%%%%%%%%%%%%%%%%%%%%%%%%%%%%%%%%%%%%%%%%

\subsubsection{Squared amplitude}

It is now straightforward to calculate the spin and color sums/averages for the squared amplitude. First, for the \emph{after} contribution, ${\cal M}^{\textrm{aft}}_{g\rightarrow q\bar{q}}$, one gets
%%%%%%%%%%%%%%%%%%%%%%%%%%%%%%%%%%%%%%%%%%%%%%%%%
\begin{eqnarray}
\frac{1}{d_A}\sum_{a_0,\, \beta_1,\, \beta_2} \frac{1}{D\!-\!2}\sum_{\lambda_0,\, h_1,\, h_2}  \Big|{\cal M}^{\textrm{aft}}_{g\rightarrow q\bar{q}}\Big|^2
&=& \frac{g^2\, T_F\, (\mu^2)^{2-\frac{D}{2}} }{\left[\left(\p_2\!-\!\frac{p_2^+}{p_1^+}\, \p_1\right)^2+\left(\frac{k_0^+}{p_1^+}\right)^2 m^2 \right]^2}\nonumber\\
&&\hspace{-3cm}\times\,
\int d^{D\!-\!2} \x_0\; \int d^{D\!-\!2} \x_{0'}\;
e^{-i(\p_1+\p_2-\k_0) \cdot \x_{00'}}\;
S^A_{00'}\; \frac{p_1^+\, p_2^+}{(k_0^+)^2}
\nonumber\\
&& \hspace{-7cm} \times \; \bigg\{ \frac{4}{D\!-\!2}\left(\frac{k_0^+}{p_1^+}\right)^2
\bigg[\left(\frac{p_1^+}{k_0^+}\right)^2+\left(\frac{p_2^+}{k_0^+}\right)^2 +\frac{D\!-\!4}{2}\bigg] \left(\p_2\!-\!\frac{p_2^+}{p_1^+}\, \p_1\right)^2
+2 \left(\frac{k_0^+}{p_1^+}\right)^4 m^2
\bigg\}
\, ,
\end{eqnarray}
%%%%%%%%%%%%%%%%%%%%%%%%%%%%%%%%%%%%%%%%%%%%%%%%%
with the adjoint dipole defined as
%%%%%%%%%%%%%%%%%%%%%%%%%%%%%%%%%%%%%%%%%%%%%%%%%
\begin{eqnarray}
S^A_{01} &\equiv & \frac{1}{d_A}\; \textrm{Tr}\big[ U_A(\x_0)  U_A^{\dagger}(\x_1)\big]\, .
\end{eqnarray}
%%%%%%%%%%%%%%%%%%%%%%%%%%%%%%%%%%%%%%%%%%%%%%%%%
Second, for the \emph{before} contribution, ${\cal M}^{\textrm{bef}}_{g\rightarrow q\bar{q}}$, one obtains
%%%%%%%%%%%%%%%%%%%%%%%%%%%%%%%%%%%%%%%%%%%%%%%%%
\begin{eqnarray}
\frac{1}{d_A}\sum_{a_0,\, \beta_1,\, \beta_2} \frac{1}{D\!-\!2}\sum_{\lambda_0,\, h_1,\, h_2}  \Big|{\cal M}^{\textrm{bef}}_{g\rightarrow q\bar{q}}\Big|^2
&=& (\mu^2)^{2-\frac{D}{2}} \, g^2\,  \; \frac{p_1^+\, p_2^+}{(k_0^+)^2}\nonumber\\
&&\hspace{-5cm}\times
\int d^{D\!-\!2} \x_1\; \int d^{D\!-\!2} \x_{1'}\; e^{-i\left(\p_1-\frac{p_1^+}{k^+_0}\k_0\right) \cdot\x_{11'}}
\int d^{D\!-\!2} \x_2\; \int d^{D\!-\!2} \x_{2'} \; e^{-i\left(\p_2-\frac{p_2^+}{k^+_0}\k_0\right) \cdot \x_{22'}}\nonumber\\
&&\hspace{-5cm}\times\;
\frac{1}{d_A}\; \textrm{Tr}\big[ U_F(\x_1) t^a  U_F^{\dagger}(\x_2)U_F(\x_{2'}) t^a  U_F^{\dagger}(\x_{1'})\big]
\nonumber\\
&& \hspace{-5cm} \times \; \bigg\{ \frac{4}{D\!-\!2}
\bigg[\left(\frac{p_1^+}{k_0^+}\right)^2+\left(\frac{p_2^+}{k_0^+}\right)^2 +\frac{D\!-\!4}{2}\bigg]\;\; {\cal B}^i_V(\x_{12},m)\; \; {\cal B}^{i\, *}_V(\x_{1'2'},m)\nonumber\\
&& \hspace{-4cm}
+2 m^2\;\; {\cal B}_S(\x_{12},m)\; \; {\cal B}^{*}_S(\x_{1'2'},m)
\bigg\}
\, .\label{bef_ampl_squared}
\end{eqnarray}

%%%%%%%%%%%%%%%%%%%%%%%%%%%%%%%%%%%%%%%%%%%%%%%%%%
The multipole appearing in eq. \eqref{bef_ampl_squared} can be rewritten as a product of fundamental dipoles, and a $N_c$-suppressed fundamental quadrupole term. In single inclusive heavy quark production, this multipole will collapse to a dipole upon integration over the transverse momentum of the un-tagged produced particle.

Finally, for the interference between the before and after contributions to the amplitude, one gets
%%%%%%%%%%%%%%%%%%%%%%%%%%%%%%%%%%%%%%%%%%%%%%%%%
\begin{eqnarray}
&&\hspace{-1.5cm}\frac{1}{d_A}\sum_{a_0,\, \beta_1,\, \beta_2} \frac{1}{D\!-\!2}\sum_{\lambda_0,\, h_1,\, h_2}  \Big[\left({\cal M}^{\textrm{aft}}_{g\rightarrow q\bar{q}}\right)^{\dagger} {\cal M}^{\textrm{bef}}_{g\rightarrow q\bar{q}}+c.c.\Big]
=- g^2\, T_F\; (\mu^2)^{2-\frac{D}{2}}\;  \frac{p_1^+\, p_2^+}{(k_0^+)^2} \nonumber\\
&&\times\;
\int d^{D\!-\!2} \x_0\;\int d^{D\!-\!2} \x_1\; e^{-i\left(\p_1-\frac{p_1^+}{k^+_0}\k_0\right) \cdot\x_{10}}\int d^{D\!-\!2} \x_2\; e^{-i\left(\p_2-\frac{p_2^+}{k^+_0}\k_0\right) \cdot \x_{20}}\nonumber\\
&&
\times \; \bigg\{ \frac{-4}{D\!-\!2}
\bigg[\left(\frac{p_1^+}{k_0^+}\right)^2+\left(\frac{p_2^+}{k_0^+}\right)^2 +\frac{D\!-\!4}{2}\bigg]\; \left(\frac{k_0^+}{p_1^+}\right)\;
\left[\p_2^i\!-\!\frac{p_2^+}{p_1^+}\, \p_1^i\right]\;{\cal B}^i_V(\x_{12},m)\nonumber\\
&&\hspace{1cm}
+2 m^2\;\left(\frac{k_0^+}{p_1^+}\right)^2
 \;{\cal B}_S(\x_{12},m)
\bigg\}\;
\frac{S_{120}}{\left[\left(\p_2\!-\!\frac{p_2^+}{p_1^+}\, \p_1\right)^2+\left(\frac{k_0^+}{p_1^+}\right)^2 m^2 \right]}\;\;
 \; + \; c.c.
\, ,\label{interference}
\end{eqnarray}
%%%%%%%%%%%%%%%%%%%%%%%%%%%%%%%%%%%%%%%%%%%%%%%%%
where we have defined the tripole operator as
%%%%%%%%%%%%%%%%%%%%%%%%%%%%%%%%%%%%%%%%%%%%%%%%%
\begin{eqnarray}
S_{120} &\equiv & \frac{1}{d_F\, C_F}\; \textrm{Tr}\big[ U_F(\x_1) t^a   U_F^{\dagger}(\x_2) t^b\big]\; U_A(\x_0)_{b\, a}
\end{eqnarray}
%%%%%%%%%%%%%%%%%%%%%%%%%%%%%%%%%%%%%%%%%%%%%%%%%
and we have used the identity $d_F\; C_F= d_A\; T_F$.

%%%%%%%%%%%%%%%%%%%%%%%%%%%%%%%%%%%%%%%%%%%%%%%%%%%%%%%%%%%%%%%%%%%%%%%%%%%%%%%%%%%%%%%%%%%%%%%%%%%%%%%
%%%%%%%%%%%%%%%%%%%%%%%%%%%%%%%%%%%%%%%%%%%%%%%%%%%%%%%%%%%%%%%%%%%%%%%%%%%%%%%%%%%%%%%%%%%%%%%%%%%%%%%
%%%%%%%%%%%%%%%%%%%%%%%%%%%%%%%%%%%%%%%%%%%%%%%%%%%%%%%%%%%%%%%%%%%%%%%%%%%%%%%%%%%%%%%%%%%%%%%%%%%%%%%
%%%%%%%%%%%%%%%%%%%%%%%%%%%%%%%%%%%%%%%%%%%%%%%%%%%%%%%%%%%%%%%%%%%%%%%%%%%%%%%%%%%%%%%%%%%%%%%%%%%%%%%

\subsubsection{Partonic cross-section for single inclusive heavy quark production}

{ The next step is to integrate over the momentum of the produced antiquark in order to obtain the single inclusive cross section at parton level, following the relation \eqref{pair_to_single_incl}.
Then, the contribution from the square of the before term reads}
%%%%%%%%%%%%%%%%%%%%%%%%%%%%%%%%%%%%%%%%%%%%%%%%%
\begin{eqnarray}
\hspace{-1cm}(2p_1^+)(2\pi)^{D\!-\!1} \frac{d\sigma^{g+A\rightarrow q+X}}{dp_1^+\, d^{D\!-\!2}\p_1} \bigg|_{\textrm{bef.-bef.}}
&=&  g^2\, T_F\;  \theta(k_0^+\!-\!p_1^+) \;  \frac{p_1^+}{k^+_0}\;
\int d^{D\!-\!2} \x_1\;\int d^{D\!-\!2} \x_{1'}\;  S^F_{11'}\; \nonumber\\
&& \hspace{-5cm}
\times\;  e^{-i\left(\p_1-\frac{p_1^+}{k^+_0}\k_0\right) \cdot\x_{11'}}\;\;
\bigg\{ \frac{4}{D\!-\!2}
\bigg[\left(\frac{p_1^+}{k_0^+}\right)^2+\left(1\!-\!\frac{p_1^+}{k_0^+}\right)^2 +\frac{D\!-\!4}{2}\bigg]\; {\cal C}_1\left(|\x_{1'1}|, m\right)\; \nonumber\\
&& %\hspace{-5cm}
%\times \;
+2 m^2\; {\cal C}_0\left(|\x_{1'1}|, m\right)
\bigg\}
%\nonumber\\
%& & \hspace{2cm}
%\times
\, ,\label{part_cross_sec_befbef}
\end{eqnarray}
%%%%%%%%%%%%%%%%%%%%%%%%%%%%%%%%%%%%%%%%%%%%%%%%%
where
%%%%%%%%%%%%%%%%%%%%%%%%%%%%%%%%%%%%%%%%%%%%%%%%%
\begin{eqnarray}
&&\hspace{-2cm}{\cal C}_1\left(|\r|, m\right)\equiv  (\mu^2)^{2-\frac{D}{2}}\; \int \frac{d^{D-2} \K}{(2\pi)^{D-2}}\; e^{-i\K \cdot \r}\; \frac{\K^2}{\Big[\K^2+m^2\Big]^2} \nonumber\\
&&\hspace{-2cm}= \frac{1}{2\pi}\, \left(\frac{m^2}{4\pi\, \mu^2}\right)^{\frac{D}{2}-2}\;
\bigg[\left(\frac{m |\r|}{2}\right)^{2-\frac{D}{2}}\; \mathrm{K}_{2-\frac{D}{2}}(m |\r|)
-\left(\frac{m |\r|}{2}\right)^{3-\frac{D}{2}}\; \mathrm{K}_{3-\frac{D}{2}}(m |\r|)
\bigg],
 \\
&&\hspace{-2cm}{\cal C}_0\left(|\r|, m\right)\equiv  (\mu^2)^{2-\frac{D}{2}}\; \int \frac{d^{D-2} \K}{(2\pi)^{D-2}}\; e^{-i\K \cdot \r}\; \frac{1}{\Big[\K^2+m^2\Big]^2}\nonumber\\
&&\hspace{-0.3cm}= \frac{1}{2\pi\, m^2}\, \left(\frac{m^2}{4\pi\, \mu^2}\right)^{\frac{D}{2}-2}\;\;\;
\left(\frac{m |\r|}{2}\right)^{3-\frac{D}{2}}\; \mathrm{K}_{3-\frac{D}{2}}(m |\r|)\, ,
\end{eqnarray}
%%%%%%%%%%%%%%%%%%%%%%%%%%%%%%%%%%%%%%%%%%%%%%%%%
with the fundamental dipole defined as
%%%%%%%%%%%%%%%%%%%%%%%%%%%%%%%%%%%%%%%%%%%%%%%%%
\begin{eqnarray}
S^F_{01} &\equiv & \frac{1}{d_F}\; \textrm{Tr}\big[ U_F(\x_0)  U_F^{\dagger}(\x_1)\big]\, .
\end{eqnarray}
%%%%%%%%%%%%%%%%%%%%%%%%%%%%%%%%%%%%%%%%%%%%%%%%%

The contribution from the square of the after term in the amplitude can be written as
%%%%%%%%%%%%%%%%%%%%%%%%%%%%%%%%%%%%%%%%%%%%%%%%%
\begin{eqnarray}
(2p_1^+)(2\pi)^{D\!-\!1} \frac{d\sigma^{g+A\rightarrow q+X}}{dp_1^+\, d^{D\!-\!2}\p_1} \bigg|_{\textrm{aft.-aft.}}
&=&  g^2\, T_F\;  \theta(k_0^+\!-\!p_1^+) \;  \frac{k^+_0}{p_1^+}\;
\int d^{D\!-\!2} \x_0\;\int d^{D\!-\!2} \x_{0'}\;  S^A_{00'}\; \nonumber\\
&& \hspace{-5cm}
\times e^{-i\left(\frac{k^+_0}{p_1^+}\p_1-\k_0\right) \cdot\x_{00'}}\;\;
\bigg\{ \frac{4}{D\!-\!2}
\bigg[\left(\frac{p_1^+}{k_0^+}\right)^2+\left(1\!-\!\frac{p_1^+}{k_0^+}\right)^2 +\frac{D\!-\!4}{2}\bigg]\; {\cal C}_1\left(|\x_{00'}|, \frac{k_0^+}{p_1^+}\, m\right)\; \nonumber\\
&& %\hspace{-5cm}
%\times \;
+2 m^2\;\left(\frac{k_0^+}{p_1^+}\right)^2
 \;{\cal C}_0\left(|\x_{00'}|, \frac{k_0^+}{p_1^+}\, m\right)
\bigg\}
%\nonumber\\
%& & \hspace{2cm}
%\times
\, .\label{part_cross_sec_aftaft}
\end{eqnarray}
%%%%%%%%%%%%%%%%%%%%%%%%%%%%%%%%%%%%%%%%%%%%%%%%%

Finally, the interference contribution reads
%%%%%%%%%%%%%%%%%%%%%%%%%%%%%%%%%%%%%%%%%%%%%%%%%
\begin{eqnarray}
&&\hspace{-1.8cm}(2p_1^+)(2\pi)^{D\!-\!1} \frac{d\sigma^{g+A\rightarrow q+X}}{dp_1^+\, d^{D\!-\!2}\p_1} \bigg|_{\textrm{interf.}}
= - g^2\, T_F\; (\mu^2)^{2-\frac{D}{2}}\; \theta(k_0^+\!-\!p_1^+) \nonumber\\
&&\times
\int d^{D\!-\!2} \x_0\;\int d^{D\!-\!2} \x_1\; \int d^{D\!-\!2} \x_2\; S_{120}\;
 e^{-i\left(\p_1-\frac{k^+_0}{p_1^+}\k_0\right) \cdot\left(\x_{12}+\frac{k^+_0}{p_1^+}\x_{20}\right)}\;\nonumber\\
&&\times
\bigg\{ \frac{-4}{D\!-\!2}
\bigg[\left(\frac{p_1^+}{k_0^+}\right)^2+\left(1\!-\!\frac{p_1^+}{k_0^+}\right)^2 +\frac{D\!-\!4}{2}\bigg]\; {\cal B}^i_V(\x_{12},m)\; {\cal B}^{i\, *}_V\left(\x_{20},\frac{k_0^+}{p_1^+} m\right)
\nonumber\\
&& \hspace{1cm}
+2 m^2\;\left(\frac{k_0^+}{p_1^+}\right)
 \;{\cal B}_S(\x_{12},m)\; {\cal B}^{*}_S\left(\x_{20},\frac{k_0^+}{p_1^+} m\right)
\bigg\}
\;\; \; + \; c.c.
\label{part_cross_sec_interf}
\end{eqnarray}
%%%%%%%%%%%%%%%%%%%%%%%%%%%%%%%%%%%%%%%%%%%%%%%%%%%%%%%%%%%%%%%%%%%%%%%%%%%%%%%%%%%%%%%%%%%%%%%%%%%%%%%
%%%%%%%%%%%%%%%%%%%%%%%%%%%%%%%%%%%%%%%%%%%%%%%%%%%%%%%%%%%%%%%%%%%%%%%%%%%%%%%%%%%%%%%%%%%%%%%%%%%%%%%

\subsection{Hadron-level cross section for heavy quark production \label{sec:hadr_level_cross_sec_real}}

Now, we want to use the results \eqref{part_cross_sec_befbef}, \eqref{part_cross_sec_aftaft} and \eqref{part_cross_sec_interf} in order to write the cross section for single inclusive production of a heavy flavored hadron in the hybrid factorization, in a high-energy dense-dilute collision. The momentum of the projectile, target and produced hadron are denoted respectively $P^{\mu}_P$, $P^{\mu}_T$ and $p^{\mu}_h$.
By choice of frame, we have $\P_P=\P_T=0$
and $P^{-}_P$ and $P^{+}_T$ are negligible, whereas the Mandelstam $s$ variable of the collision is given by $s\simeq 2 P^{+}_P\; P^{-}_T$. The Feynman $x_F$ variable is defined by $x_F\equiv p^{+}_h/P^{+}_P$.

Neglecting for the moment possible contributions from the eventual intrinsic heavy flavor content of the projectile, and from heavy quark production during jet fragmentation, the picture is the following: A large-$x_B$ gluon with momentum $k_0^{\mu}$ is picked inside the projectile, then it collide on the target, producing a heavy quark of momentum $p_1^{\mu}$, according to the partonic cross-section sum of \eqref{part_cross_sec_befbef}, \eqref{part_cross_sec_aftaft} and \eqref{part_cross_sec_interf}, and, finally, the heavy quark  of momentum $p_1^{\mu}$ fragments into heavy flavored hadron $h$ of momentum $p^{\mu}_h$.

So, the hadronic cross section is obtained from the partonic cross section as
%%%%%%%%%%%%%%%%%%%%%%%%%%%%%%%%%%%%%%%%%%%%%%%%%
\begin{eqnarray}
(2p_h^+)(2\pi)^{D\!-\!1} \frac{d\sigma^{p+A\rightarrow h+X}}{dp_h^+\, d^{D\!-\!2}\p_h}
&=& \int_{0}^{1} d x_B\;  g^0(x_B)  \int_{0}^{1}\frac{d\zeta}{\zeta^{D-2}}\; D^0_{h/q}(\zeta)\nonumber\\
&&\times\;\;\;
(2p_1^+)(2\pi)^{D\!-\!1} \frac{d\sigma^{g+A\rightarrow q+X}}{dp_1^+\, d^{D\!-\!2}\p_1}
\, ,\label{hadr_cross_sec_def}
\end{eqnarray}
%%%%%%%%%%%%%%%%%%%%%%%%%%%%%%%%%%%%%%%%%%%%%%%%%
where, due to the collinear approximation, one has
%%%%%%%%%%%%%%%%%%%%%%%%%%%%%%%%%%%%%%%%%%%%%%%%%
\begin{eqnarray}
k_0^+= x_B\, P^{+}_P\; , \quad \k_0=0\; , \quad p_1^+=\frac{p_h^+}{\zeta}\;\; \textrm{and} \quad \p_1=\frac{\p_h}{\zeta}\, .
\end{eqnarray}
%%%%%%%%%%%%%%%%%%%%%%%%%%%%%%%%%%%%%%%%%%%%%%%%%
Since we are interested only in the leading-order result for this channel, we can replace the bare PDFs and FFs by the renormalized ones.

All in all, one obtains from  equations \eqref{part_cross_sec_befbef}, \eqref{part_cross_sec_aftaft} and \eqref{part_cross_sec_interf} the following three contributions to the hadronic cross section:
%%%%%%%%%%%%%%%%%%%%%%%%%%%%%%%%%%%%%%%%%%%%%%%%%
\begin{eqnarray}
\hspace{-1cm}(2\pi)^{D\!-\!2}\, \frac{d\sigma^{p+A\rightarrow h+X}}{d x_F\, d^{D\!-\!2}\p_h}\bigg|_{\textrm{bef.-bef.}}
&=& \int_{x_F}^{1} d x_B\;  g(x_B,\mu^2)  \int_{\frac{x_F}{x_B}}^{1}
\frac{d\zeta}{\zeta^{D-2}}\; D_{h/q}(\zeta,\mu^2)\;\;\;
 \alpha_s\, T_F\;    \frac{x_F}{x_B\, \zeta}\nonumber\\
&&
\times\;\int d^{D\!-\!2} \x_1\;\int d^{D\!-\!2} \x_{2}\;  S^F_{12}\;
e^{-\frac{i}{\zeta}\, \p_h \cdot\x_{12}}\nonumber\\
&&\hspace{-5cm}\times
\bigg\{ \frac{4}{D\!-\!2}
\bigg[\left(\frac{x_F}{x_B\, \zeta}\right)^2+\left(1\!-\!\frac{x_F}{x_B\, \zeta}\right)^2 +\frac{D\!-\!4}{2}\bigg]\; {\cal C}_1\left(|\x_{12}|, m\right)\;
+2 m^2\; {\cal C}_0\left(|\x_{12}|, m\right)
\bigg\}
\, ,\label{hadr_cross_sec_befbef}
\end{eqnarray}
%%%%%%%%%%%%%%%%%%%%%%%%%%%%%%%%%%%%%%%%%%%%%%%%%
%%%%%%%%%%%%%%%%%%%%%%%%%%%%%%%%%%%%%%%%%%%%%%%%%
\begin{eqnarray}
 (2\pi)^{D\!-\!2}\, \frac{d\sigma^{p+A\rightarrow h+X}}{d x_F\, d^{D\!-\!2}\p_h}\bigg|_{\textrm{aft.-aft.}}
&=& \int_{x_F}^{1} d x_B\;  g(x_B,\mu^2)  \int_{\frac{x_F}{x_B}}^{1}\frac{d\zeta}{\zeta^{D-2}}\; D_{h/q}(\zeta,\mu^2)\;\;\;
 \alpha_s\, T_F\;    \frac{x_B\, \zeta}{x_F}\nonumber\\
 &&\times
\int d^{D\!-\!2} \x_1\;\int d^{D\!-\!2} \x_{2}\;  S^A_{12} \;   e^{-i\frac{x_B}{x_F}\, \p_h \cdot\x_{12}}
\nonumber\\
&& \hspace{-5cm}
\times \;
\bigg\{ \frac{4}{D\!-\!2}
\bigg[\left(\frac{x_F}{x_B\, \zeta}\right)^2+\left(1\!-\!\frac{x_F}{x_B\, \zeta}\right)^2 +\frac{D\!-\!4}{2}\bigg]\; {\cal C}_1\left(|\x_{12}|, \frac{x_B\, \zeta}{x_F}\, m\right)\;
%\nonumber\\
%&& %\hspace{-5cm}
%\times \;
\label{hadr_cross_sec_aftaft}\\&&\hspace{2cm}
+2 m^2\; \left(\frac{x_B\, \zeta}{x_F}\right)^2 {\cal C}_0\left(|\x_{12}|, \frac{x_B\, \zeta}{x_F}\, m\right)
\bigg\}
\nonumber
\end{eqnarray}
%%%%%%%%%%%%%%%%%%%%%%%%%%%%%%%%%%%%%%%%%%%%%%%%%
and
%%%%%%%%%%%%%%%%%%%%%%%%%%%%%%%%%%%%%%%%%%%%%%%%%
\begin{eqnarray}
 (2\pi)^{D\!-\!2}\, \frac{d\sigma^{p+A\rightarrow h+X}}{d x_F\, d^{D\!-\!2}\p_h}\bigg|_{\textrm{interf.}}
&=& \int_{x_F}^{1} d x_B\;  g(x_B,\mu^2)  \int_{\frac{x_F}{x_B}}^{1}\frac{d\zeta}{\zeta^{D-2}}\; D_{h/q}(\zeta,\mu^2)\;
 (-1)\, \alpha_s\, T_F\;  \left(\mu^2\right)^{2-\frac{D}{2}}
 \nonumber\\&&\times
  \int d^{D\!-\!2} \x_0\;
\int d^{D\!-\!2} \x_1\;
%\nonumber\\
%&& \hspace{-5cm}
%\times
 \int d^{D\!-\!2} \x_{2}\;  S_{120}\; e^{-\frac{i}{\zeta}\, \p_h \cdot \left[\x_{12}-\frac{x_B\, \zeta}{x_F}\, \x_{02}\right]}\;\;
\nonumber\\&&
\hspace{-5cm}
\times
\bigg\{ \frac{- 4}{D\!-\!2}
\bigg[\left(\frac{x_F}{x_B\, \zeta}\right)^2+\left(1\!-\!\frac{x_F}{x_B\, \zeta}\right)^2 +\frac{D\!-\!4}{2}\bigg]\;
%{\color{red} \left(\frac{x_B}{x_F}\right)^2  }
{\cal B}^i_V(\x_{12},m)\; {\cal B}^{i\, *}_V\left(\x_{20},\frac{x_B\, \zeta}{x_F}\, m\right)\nonumber\\
&& \hspace{-1.5cm}
%\times \;
+2 m^2\;\left(\frac{x_B\, \zeta}{x_F}\right)
 \;{\cal B}_S(\x_{12},m)\; {\cal B}^{*}_S\left(\x_{20},\frac{x_B\, \zeta}{x_F}\, m\right)
\bigg\}\;\; \; + \; c.c.
\label{hadr_cross_sec_interf}
\end{eqnarray}
%%%%%%%%%%%%%%%%%%%%%%%%%%%%%%%%%%%%%%%%%%%%%%%%%
%%%%%%%%%%%%%%%%%%%%%%%%%%%%%%%%%%%%%%%%%%%%%%%%%%
%%%%%%%%%%%%%%%%%%%%%%%%%%%%%%%%%%%%%%%%%%%%%%%%%
\subsection{Large mass limit of the hadronic cross section}
We also consider the large mass limit of the hadronic cross section. The before-before  and after-after contributions to the total hadronic cross section are written in terms of the functions  ${\cal C}_0(|\r|,m)$ and ${\cal C}_1(|\r|,m)$. These functions, when expanded in the large mass limit, can be approximated as
\begin{eqnarray}
&&C_0(|\r|,m)
%\simeq  \left(\mu^2\right)^{2-\frac{D}{2}}\frac{1}{m^4}\,  \int\frac{d^{D-2}\K}{(2\pi)^{D-2}}\, e^{-i\K\cdot\r}
\simeq \left(\mu^2\right)^{2-\frac{D}{2}}\frac{1}{m^4} \, \left( 1+\frac{2}{m^2}\partial^2_{\r}   \right) \, \delta^{(D-2)}(\r)\; ,\\
&&C_1(|\r|,m)
%\simeq \left(\mu^2\right)^{2-\frac{D}{2}}\frac{1}{m^4}\, \left( -\partial_{\r}^2 \right) \int\frac{d^{D-2}\K}{(2\pi)^{D-2}}\, e^{-i\K\cdot\r}
\simeq \left(\mu^2\right)^{2-\frac{D}{2}}\frac{1}{m^4}\, \left( -\, \partial_{\r}^2 \right)\, \delta^{(D-2)}(\r)\; .
\end{eqnarray}
The interference contribution to the cross section is written in terms functions ${\cal B}^i_V(\r, m)$ and ${\cal B}_S(\r, m)$, whose leading term in the large mass limit reads
\begin{eqnarray}
&&{\cal B}^i_V(\r,m)
%\simeq \frac{1}{m^2} \, \left( -i\partial^i_{\r} \right) \int \frac{d^{D-2}\K}{(2\pi)^{D-2}} e^{i\K\cdot\r}
\simeq \frac{1}{m^2} \, \left( - \, i \, \partial_r^{i} \right)\, \delta^{(D-2)}(\r) \; ,\\
&&{\cal B}_S(\r,m)
%\simeq \frac{1}{m^2} \, \int \frac{d^{D-2}\K}{(2\pi)^{D-2}} e^{i\K\cdot\r}
\simeq \frac{1}{m^2} \, \left( 1+\frac{1}{m^2}\partial^2_{\r} \right) \delta^{(D-2)}(\r)\; .
\end{eqnarray}
Using these equations, we can obtain the large mass limit of the each contribution to the hadronic cross section that reads
\begin{eqnarray}
\hspace{-1cm}
&&(2\pi)^{D\!-\!2}\, \frac{d\sigma^{p+A\rightarrow h+X}}{d x_F\, d^{D\!-\!2}\p_h}\bigg|_{\textrm{bef.-bef.}}
= \int_{x_F}^{1} d x_B\;  g(x_B,\mu^2)  \int_{\frac{x_F}{x_B}}^{1}
\frac{d\zeta}{\zeta^{D-2}}\; D_{h/q}(\zeta,\mu^2)\;\;\;
 \alpha_s\, T_F\;    \frac{x_F}{x_B\, \zeta}\nonumber\\
&&\times \, \left( \mu^2 \right)^{2-\frac{D}{2}}
\int d^{D-2}\x_1 \int d^{D-2}\x_2 \; \delta^{(D-2)}(\x_{12})\nonumber\\
&&\times
 \Bigg\{ \frac{2}{m^2}-\frac{1}{m^4}\frac{\p_h^2}{\zeta^2}
\left[4-
\frac{4}{D-2}\left[ \left( \frac{x_F}{x_B\, \zeta} \right)^2 + \left( 1-\frac{x_F}{x_B\, \zeta} \right)^2 + \frac{D-4}{2} \right]
\right]
\nonumber\\
&&
%\hspace{3cm}
+\,
\frac{1}{m^4}
\left[4-
\frac{4}{D-2}\left[ \left( \frac{x_F}{x_B\, \zeta} \right)^2 + \left( 1-\frac{x_F}{x_B\, \zeta} \right)^2+ \frac{D-4}{2} \right]
\right]
%\nonumber\\
%&&
%\hspace{3cm}
%\times
\Bigg[ \left(\partial^2_{\x_{1}}S^F_{12}\right) -2i\frac{1}{\zeta}\p_h^i\left( \partial^i_{\x_{1}}S^F_{12}\right) \Bigg]\nonumber\\
&&
+{\cal O}\left( \frac{1}{m^6} \right)
\Bigg\},\nonumber
\end{eqnarray}
\begin{eqnarray}
&& (2\pi)^{D\!-\!2}\, \frac{d\sigma^{p+A\rightarrow h+X}}{d x_F\, d^{D\!-\!2}\p_h}\bigg|_{\textrm{aft.-aft.}}
= \int_{x_F}^{1} d x_B\;  g(x_B,\mu^2)  \int_{\frac{x_F}{x_B}}^{1}\frac{d\zeta}{\zeta^{D-2}}\; D_{h/q}(\zeta,\mu^2)\;\;\;
 \alpha_s\, T_F\;    \frac{x_F}{x_B\, \zeta}\nonumber\\
&&\times \, \left( \mu^2 \right)^{2-\frac{D}{2}}
\int d^{D-2}\x_1 \int d^{D-2}\x_2 \; \delta^{(D-2)}(\x_{12})\nonumber\\
&&
\times
\Bigg\{ \frac{2}{m^2}
-\frac{1}{m^4}
\frac{\p_h^2}{\zeta^2}
\left[4-
\frac{4}{D-2}\left[ \left( \frac{x_F}{x_B\, \zeta} \right)^2 + \left( 1-\frac{x_F}{x_B\, \zeta} \right) ^2+ \frac{D-4}{2} \right]
\right]
\nonumber\\
&&
%\hspace{3cm}
+\frac{1}{m^4}
\left( \frac{x_F}{x_B\, \zeta} \right)^2
\left[4-
\frac{4}{D-2}
\left[ \left( \frac{x_F}{x_B\, \zeta} \right)^2 + \left( 1-\frac{x_F}{x_B\, \zeta} \right)^2 + \frac{D-4}{2} \right]
\right]
\nonumber\\
%&&
%\hspace{3cm}
%\times\, \int d^{D-2}x_1
&& \hskip 2.3cm \times \Bigg[ \left(\partial^2_{\x_{1}}S^A_{12}\right) - 2i\frac{x_B}{x_F}\p_h^i\left( \partial^i_{\x_{1}}S^A_{12}\right) \Bigg]
+{\cal O}\left( \frac{1}{m^6} \right)
\Bigg\},
\end{eqnarray}
\begin{eqnarray}
&& (2\pi)^{D\!-\!2}\, \frac{d\sigma^{p+A\rightarrow h+X}}{d x_F\, d^{D\!-\!2}\p_h}
\bigg|_{\textrm{interf.}}
= \int_{x_F}^{1} d x_B\;  g(x_B,\mu^2)  \int_{\frac{x_F}{x_B}}^{1}\frac{d\zeta}{\zeta^{D-2}}\; D_{h/q}(\zeta,\mu^2)\;
 (-1)\, \alpha_s\, T_F\;   \nonumber\\
&&\times \, \left( \frac{x_F}{x_B\, \zeta} \right) \left( \mu^2 \right)^{ 2-\frac{D}{2}}
\int d^{D-2}\x_1 \int d^{D-2}\x_2  \int d^{D-2}\x_0 \; \delta^{(D-2)}(\x_{12})\;  \delta^{(D-2)}(\x_{02}) \nonumber\\
 &&\times
 \Bigg\{
\frac{2}{m^2}
-\frac{1}{m^4}\frac{\p_h^2}{\zeta^2}
\left[4-
\frac{4}{D-2}\left[ \left( \frac{x_F}{x_B\, \zeta} \right)^2 + \left( 1-\frac{x_F}{x_B\, \zeta} \right)^2 + \frac{D-4}{2} \right]
\right]
\nonumber\\
&&
%\hspace{3cm}
+\frac{1}{m^4} \left( \frac{x_F}{x_B\, \zeta} \right) \frac{4}{D-2} \left[ \left( \frac{x_F}{x_B\, \zeta} \right)^2 + \left( 1-\frac{x_F}{x_B\, \zeta} \right)^2 + \frac{D-4}{2} \right]
\nonumber\\
&&\times \,
 \left[ \left( \partial^i_{\x_{1}} \partial^i_{\x_{0}} S_{120}\right)-\frac{i}{\zeta}\p_h^i\left(\partial^i_{\x_{0}}S_{120}\right)-\frac{i}{\zeta}\p_h^i\left(\frac{x_B\, \zeta}{x_F}\right)\left(\partial^i_{\x_{1}}S_{120}\right) \right]
\nonumber\\
&& +\frac{2}{m^4}
\left[ \left( \partial^2_{\x_{1}}S_{120} \right)
-2\frac{i}{\zeta}\p_h^i\left( \partial^{i}_{\x_{1}} S_{120}\right)+\left( \frac{x_F}{x_B\, \zeta}\right)^2 \left( \partial^2_{\x_{0}}S_{120}\right)-2\frac{i}{\zeta}\p_h^i\left(\frac{x_F}{x_B\, \zeta}\right) \left( \partial^i_{\x_{0}}S_{120}\right)
\right]\nonumber\\
&&+{\cal O}\left( \frac{1}{m^6}\right)
\Bigg\}+c.c.
 \end{eqnarray}
Each contribution to the total hadron-level cross section can be simplified. Any term with a single transverse derivative acting on an adjoint or fundamental dipole or the tripole operator can be dropped. This is due to the fact that each transverse derivative brings a generator of the $SU(N_c)$ group either in the fundamental or adjoint representation inside the trace. These terms become a trace of single generator by realising the delta functions, hence they vanish. Moreover, the delta functions also leads to simplifications on the tripole operator. It can be reduced to either the identity or a fundamental or adjoint dipole. Specifically, we have
\begin{equation}
S_{222}=\mathds{1}, \hspace{1cm} S_{220}= S^A_{20}, \hspace{1cm} S_{122}=S^F_{12}\,.
\end{equation}
After all these simplifications, the total hadron-level cross section in the large mass limit reads
\begin{eqnarray}
&& (2\pi)^{D\!-\!2}\, \frac{d\sigma^{p+A\rightarrow h+X}}{d x_F\, d^{D\!-\!2}\p_h}
= \int_{x_F}^{1} d x_B\;  g(x_B,\mu^2)  \int_{\frac{x_F}{x_B}}^{1}\frac{d\zeta}{\zeta^{D-2}}\; D_{h/q}(\zeta,\mu^2)\;
\alpha_s\, T_F  \left( \frac{x_F}{x_B\, \zeta} \right)
\left( \mu^2 \right)^{ 2-\frac{D}{2}}
\nonumber\\
&&
\hspace{3cm}
\times\, \frac{1}{m^4}
\int d^{D-2}\x_1 \int d^{D-2}\x_2  \int d^{D-2}\x_0 \; \delta^{(D-2)}(\x_{12})\;  \delta^{(D-2)}(\x_{02})
 \nonumber\\
 &&
\hspace{3cm}
 \times
  \frac{-4}{D-2} \left[ \left( \frac{x_F}{x_B\, \zeta} \right)^2 + \left( 1-\frac{x_F}{x_B\, \zeta} \right)^2 + \frac{D-4}{2} \right]
 \nonumber\\
 &&
 \hspace{1cm}
 \times
 \left[
 \partial^2_{\x_1}S^F_{12}+\left( \frac{x_F}{x_B\, \zeta} \right)^2\partial^2_{\x_0}S^A_{02}
 +\left( \frac{x_F}{x_B\, \zeta} \right) \partial^i_{\x_1} \partial^i_{\x_0}\left[S_{120}+S_{210}\right]
 \right]+ {\cal O}\left( \frac{1}{m^6} \right)
 \label{LM_S}
 \end{eqnarray}
It is straightforward to calculate the action of the transverse derivatives on the dipole and tripole operators:
\begin{eqnarray}
&&
\hspace{-1cm}
\partial^2_{\x_1}S^F_{12}\bigg|_{\x_2\to\x_1}=-2\, \frac{C_F}{d_A}\, g^2\int_{-\infty}^{+\infty}dx^+\int_{-\infty}^{x^+}dz^+
\left[ \partial^i_{\x_1}{\cal A}^-_a(x^+,\x_1)\right]
\nonumber\\
&&\hspace{6.5cm}
\times
\,U_A(x^+,z^+;\x_1)^{ab} \left[ \partial^i_{\x_1}{\cal A}^-_b(z^+,\x_1)\right]\, ,\label{LapF}
\\
&&
\hspace{-1cm}
\partial^2_{\x_0}S^A_{02}\bigg|_{\x_2\to\x_0}=-2\,\frac{C_A}{d_A}\,g^2\int_{-\infty}^{+\infty}dx^+\int_{-\infty}^{x^+}dz^+
\left[ \partial^i_{\x_0}{\cal A}^-_a(x^+,\x_0)\right]
\nonumber\\
&&\hspace{6.5cm}
\times
\,U_A(x^+,z^+;\x_0)^{ab} \left[ \partial^i_{\x_0}{\cal A}^-_b(z^+,\x_0)\right]\, ,
\label{LapA}
\\
&&
\hspace{-1cm}
\partial^i_{\x_0}\partial^i_{\x_1}
S^F_{120}\bigg|_{\x_1\to\x_2; \, \x_0\to\x_2}= \frac{C_A}{d_A}\, g^2\int_{-\infty}^{+\infty}dx^+\int_{-\infty}^{x^+}dz^+
\left[ \partial^i_{\x_2}{\cal A}^-_a(x^+,\x_2)\right]
\nonumber\\
&&\hspace{6.5cm}
\times
\,U_A(x^+,z^+;\x_2)^{ab} \left[ \partial^i_{\x_2}{\cal A}^-_b(z^+,\x_2)\right]\, .
\label{DerTripole}
\end{eqnarray}
Plugging \eqref{LapF}, \eqref{LapA} and \eqref{DerTripole} in the total hadron-level cross section, \eqref{LM_S}, we get
\begin{eqnarray}
&& (2\pi)^{D\!-\!2}\, \frac{d\sigma^{p+A\rightarrow h+X}}{d x_F\, d^{D\!-\!2}\p_h}
= \int_{x_F}^{1} d x_B\;  g(x_B,\mu^2)  \int_{\frac{x_F}{x_B}}^{1}\frac{d\zeta}{\zeta^{D-2}}\; D_{h/q}(\zeta,\mu^2)\;
\frac{\alpha_s\, T_F}{m^4}  \left( \frac{x_F}{x_B\, \zeta} \right)
\left( \mu^2 \right)^{ 2-\frac{D}{2}}
\nonumber\\
&&
%\hspace{3cm}
\times\,
\frac{2}{D-2}\left[
\left(\frac{x_F}{x_B\, \zeta}\right)^2+\left( 1-\frac{x_F}{x_B\, \zeta}\right)^2+\frac{D-4}{2}
 \right]
 \left[
\left(\frac{x_F}{x_B\, \zeta}\right)^2+\left( 1-\frac{x_F}{x_B\, \zeta}\right)^2+\frac{2C_F-C_A}{C_A}
  \right]
  \nonumber\\
  &&
  \times\,
  \frac{C_A}{C_F}\int d^{D-2}\x_1 \int d^{D-2}\x_2 \, \delta^{(D-2)}(\x_{12}) (-1) \left( \partial^2_{\x_1}S^F_{12}\right) +{\cal O}\left( \frac{1}{m^6} \right).
  \label{hqlimitfinal}
 \end{eqnarray}
It is possible to further simplify the expression of the cross section by adopting a model for the fundamental dipole operator. Then, this model can be used to explicitly perform the transverse integrations. For instance, in the Golec-Biernat-W\"usthoff (GBW) model \cite{GBW} (or in the McLerran-Venugopalan (MV) model \cite{MV} neglecting the logarithm in the exponent)  the fundamental dipole is written as
\begin{equation}
S^F_{12}=e^{-\frac{\x_{12}^2Q_s^2}{4}}.
\end{equation}
Thus, the cross section in the GBW model reads
\begin{eqnarray}
&&
\hspace{-0.3cm}
(2\pi)^{D\!-\!2}\, \frac{d\sigma^{p+A\rightarrow h+X}}{d x_F\, d^{D\!-\!2}\p_h}
= \int_{x_F}^{1} d x_B\;  g(x_B,\mu^2)  \int_{\frac{x_F}{x_B}}^{1}\frac{d\zeta}{\zeta^{D-2}}\; D_{h/q}(\zeta,\mu^2)\;
\frac{\alpha_s\, T_F}{m^4}  \left( \frac{x_F}{x_B\, \zeta} \right)
\left( \mu^2 \right)^{ 2-\frac{D}{2}}
\nonumber
\\
&&
\hspace{4.5cm}
%\hspace{3cm}
\times
%\frac{2}{D-2}
\left[
\left(\frac{x_F}{x_B\, \zeta}\right)^2+\left( 1-\frac{x_F}{x_B\, \zeta}\right)^2+\frac{D-4}{2}
 \right]
 \nonumber\\
 &&
\hspace{2cm}
 \times\,
 \left[
\left(\frac{x_F}{x_B\, \zeta}\right)^2+\left( 1-\frac{x_F}{x_B\, \zeta}\right)^2+\frac{2C_F-C_A}{C_A}
  \right]
  %\nonumber\\
  %&&
  %\times\,
 \frac{C_A}{C_F}\, Q_s^2\, S_{\perp} + {\cal O}\left( \frac{1}{m^6} \right)\; .
 % \frac{C_A}{C_F}\int d^{D-2}\x_1 \int d^{D-2}\x_2 \, \delta^{(D-2)}(\x_{12}) (-1) \left( \partial^2_{\x_1}S^F_{12}\right)
\end{eqnarray}
Here $S_\perp$ is the transverse area of the target, { introduced to replace the $\x_1$ integration, as required due to the impact parameter independence of the GBW model.}

\subsection{About the intrinsic heavy flavor contribution \label{sec:intrinsic_charm}}

So far, we have considered only the extrinsic contribution to heavy flavored hadron production, where the heavy quarks are pair-produced perturbatively upon scattering on the target of an incoming gluon from the projectile.
This is indeed expected to be usually the dominant contribution to heavy flavored hadron production, but maybe not the only one.

Another sizable contribution might come from the intrinsic heavy flavor content of the proton projectile at non-perturbative level, before any perturbative evolution. Such intrinsic heavy flavor contribution is known to be power-suppressed in the large mass limit. But, still, it might be relevant in the case of the charm quark and possibly also (but to a lesser extent) in the case of the bottom quark.

In a fixed flavor number scheme (FFNS) like the one we are using, where the heavy flavors are not considered active, the intrinsic charm and bottom contributions can be taken into account by providing charm and bottom PDFs inside the proton, and considering the relevant diagrams with an incoming charm or bottom quark. However, note that in this scheme, these heavy flavor PDF are independent of the factorization scale, and thus are not affected by perturbative evolution, as explained for example in Refs. \cite{Ball:2015tna,Ball:2015dpa}.

Then, the leading-order term for the intrinsic heavy flavor contribution to heavy flavored hadron production is identical to the quark channel leading-order term for unidentified hadron (or pion) production, up to the replacement of PDF and FF. One has
%%%%%%%%%%%%%%%%%%%%%%%%%%%%%%%%%%%%%%%%%%%%%%%%%
\begin{eqnarray}
& & (2\pi)^{D\!-\!2}\, \frac{d\sigma^{p+A\rightarrow h+X}}{d x_F\, d^{D\!-\!2}\p_h}\bigg|_{\textrm{intr. heavy flavor}}
= \int_{0}^{1} d x_B\;  Q(x_B)  \int_{0}^{1}\frac{d\zeta}{\zeta^{D-2}}\; D_{h/q}(\zeta,\mu^2)\;\;\;
 \nonumber\\
 && \hspace{1cm} \times
  x_B\, \delta\left(x_B\!-\!\frac{x_F}{\zeta}\right)
\int d^{D\!-\!2} \x_1\;\int d^{D\!-\!2} \x_{2}\;  S^F_{12} \;   e^{-\frac{i}{\zeta}\, \p_h \cdot\x_{12}}
\; ,\label{hadr_cross_sec_intr_charm_LO}
\end{eqnarray}
%%%%%%%%%%%%%%%%%%%%%%%%%%%%%%%%%%%%%%%%%%%%%%%%%
where $Q(x_B)$ is the scale-invariant PDF for the intrinsic charm or bottom content of the proton projectile. It is a non-perturbative input which has to be modeled (see for example
Refs. \cite{Brodsky:1980pb,Blumlein:2015qcn}) and/or fitted on experimental data, like the initial condition at low $\mu$ for the PDFs of massless partons.

Formally, the intrinsic heavy flavor contribution \eqref{hadr_cross_sec_intr_charm_LO} is of order ${\cal O}(1)$ in perturbation theory, whereas the extrinsic contribution starts at order ${\cal O}(\alpha_s)$, see eqs. \eqref{hadr_cross_sec_befbef}, \eqref{hadr_cross_sec_aftaft} and \eqref{hadr_cross_sec_interf}.
Hence, from a formal point of view, if one wants to include both the extrinsic and intrinsic contributions, one should also calculate and include the ${\cal O}(\alpha_s)$ corrections to the intrinsic contribution \eqref{hadr_cross_sec_intr_charm_LO}.
However, $Q(x_B)$ is expected to be much smaller than the gluon PDF, so that the the difference in perturbative order can be overcome.

In practice, one can estimate the leading-order term for each contribution in the kinematical range of interest thanks to the formulae \eqref{hadr_cross_sec_befbef}, \eqref{hadr_cross_sec_aftaft} and \eqref{hadr_cross_sec_interf}, and \eqref{hadr_cross_sec_intr_charm_LO}, and start to worry about the ${\cal O}(\alpha_s)$ corrections to the intrinsic contribution only
when its ${\cal O}(1)$ term is non-negligible compared to the extrinsic contribution.
The calculation of these ${\cal O}(\alpha_s)$ corrections to the intrinsic contribution can be done with the same method used in this paper. However, this is beyond the scope of this study, which focuses on the extrinsic contribution.

%%%%%%%%%%%%%%%%%%%%%%%%%%%%%%%%%%%%%%%%%%%%%%%%%%%%%%%%%%%%%%%%%%%%%%%%%%%%%%%%%%%%%%%%%%%%%%%%%%%%%%%
%%%%%%%%%%%%%%%%%%%%%%%%%%%%%%%%%%%%%%%%%%%%%%%%%%%%%%%%%%%%%%%%%%%%%%%%%%%%%%%%%%%%%%%%%%%%%%%%%%%%%%%
%%%%%%%%%%%%%%%%%%%%%%%%%%%%%%%%%%%%%%%%%%%%%%%%%%%%%%%%%%%%%%%%%%%%%%%%%%%%%%%%%%%%%%%%%%%%%%%%%%%%%%%
%%%%%%%%%%%%%%%%%%%%%%%%%%%%%%%%%%%%%%%%%%%%%%%%%%%%%%%%%%%%%%%%%%%%%%%%%%%%%%%%%%%%%%%%%%%%%%%%%%%%%%%

%%%%%%%%%%%%%%%%%%%%%%%%%%%%%%%%%%%%%%%%%%%%%%%%%%%%%%%%%%%%%%%%%%%%%%%%%%%%%%%%%%%%%%%%%%%%%%%%%%%%%%%
%%%%%%%%%%%%%%%%%%%%%%%%%%%%%%%%%%%%%%%%%%%%%%%%%%%%%%%%%%%%%%%%%%%%%%%%%%%%%%%%%%%%%%%%%%%%%%%%%%%%%%%
%%%%%%%%%%%%%%%%%%%%%%%%%%%%%%%%%%%%%%%%%%%%%%%%%%%%%%%%%%%%%%%%%%%%%%%%%%%%%%%%%%%%%%%%%%%%%%%%%%%%%%%
%%%%%%%%%%%%%%%%%%%%%%%%%%%%%%%%%%%%%%%%%%%%%%%%%%%%%%%%%%%%%%%%%%%%%%%%%%%%%%%%%%%%%%%%%%%%%%%%%%%%%%%

\section{Heavy quark loop correction to the gluon in the gluon scattering amplitude on the background field}
\label{sec4}

%%%%%%%%%%%%%%%%%%%%%%%%%%%%%%%%%%%%%%%%%%%%%%%%%%%%%%%%%%%%%%%%%%%%%%%%%%%%%%%%%%%%%%%%%%%%%%%%%%%%%%%
%%%%%%%%%%%%%%%%%%%%%%%%%%%%%%%%%%%%%%%%%%%%%%%%%%%%%%%%%%%%%%%%%%%%%%%%%%%%%%%%%%%%%%%%%%%%%%%%%%%%%%%
%%%%%%%%%%%%%%%%%%%%%%%%%%%%%%%%%%%%%%%%%%%%%%%%%%%%%%%%%%%%%%%%%%%%%%%%%%%%%%%%%%%%%%%%%%%%%%%%%%%%%%%
%%%%%%%%%%%%%%%%%%%%%%%%%%%%%%%%%%%%%%%%%%%%%%%%%%%%%%%%%%%%%%%%%%%%%%%%%%%%%%%%%%%%%%%%%%%%%%%%%%%%%%%

\subsection{Final state heavy quark pair contribution to gluon merging}

\subsubsection{Momentum space}

\begin{figure}
\setbox1\hbox to 10cm{
%\fcolorbox{white}{white}{
 \includegraphics{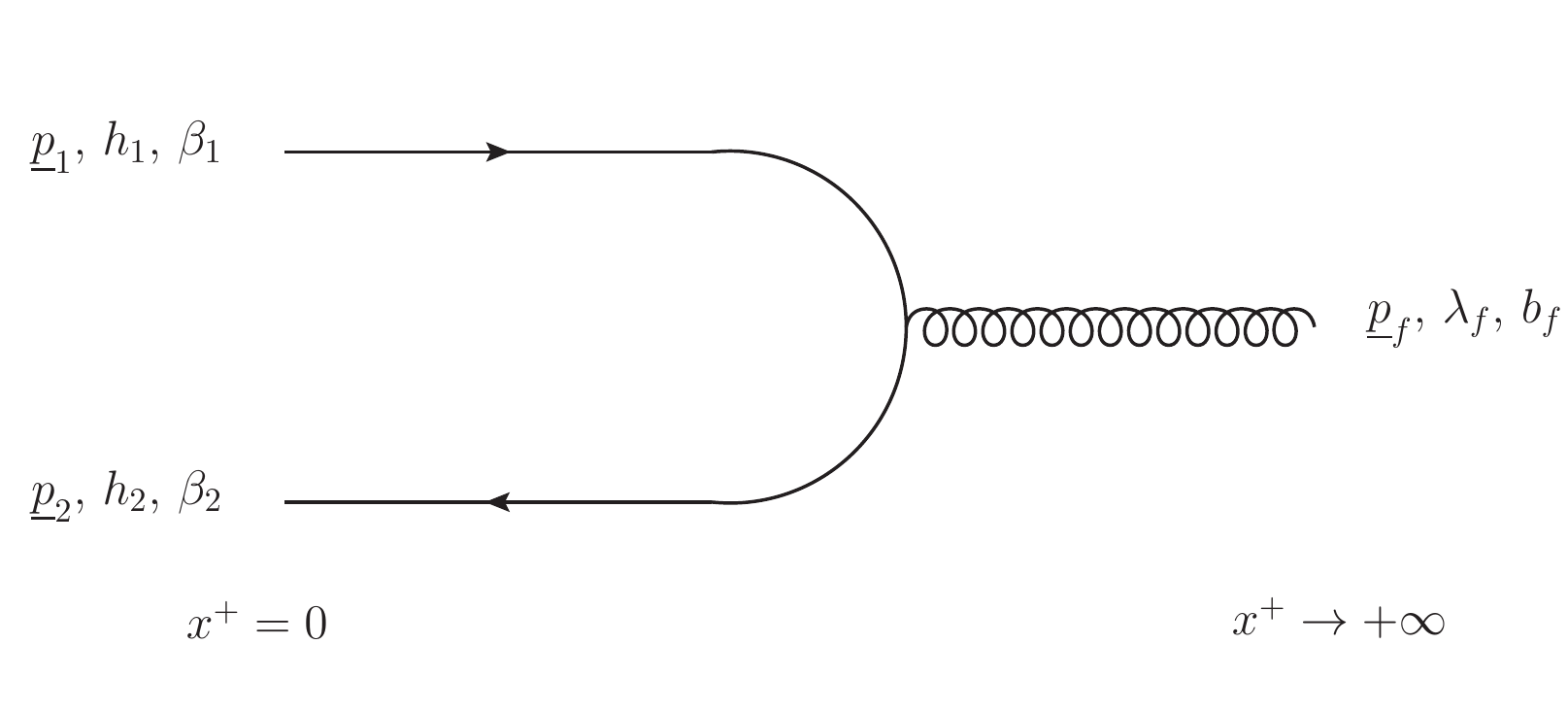}
%}
}
\begin{center}
\hspace{-4cm}
\resizebox*{6cm}{!}{\box1}
\caption{\label{Fig:qqbar2g_FS}Tree-level contribution to the $q\bar{q}$ Fock component of the outgoing $g$ state.}
\end{center}
\end{figure}

The Fock state decomposition of the one-gluon final state can be calculated directly following the rules of light-front perturbation theory. But, alternatively, it can also be obtained by taking the conjugate of the Fock state decomposition of the one-gluon initial state, Eq. \eqref{phys_gluon_decomp_mom}. In both cases, one obtains
%%%%%%%%%%%%%%%%%%%%%%%%%%%%%%%%%%%%%%%%%%%%%%%%%
\begin{eqnarray}
\langle g(\underline{p_f},\lambda_f, b_f)_{\textrm{phys}}|
&=& \sqrt{Z_{A}} \bigg[\langle 0|\, a(\underline{p_f},\lambda_f, b_f)
\nonumber\\
& &
 \hspace{-0.8cm}
+\sum_{q \bar{q}\textrm{ states}}  \left(\Psi_{q_1 \bar{q}_2}^{g_0}\right)^{\dagger}\;\;  (t^{b_f})_{\beta_2\, \beta_1}\; \;
\langle 0|\,
d(\underline{p_2},h_2,\beta_2)\,
b(\underline{p_1},h_1,\beta_1) \label{phys_gFS_decomp_mom}
\\
&&
\hspace{-0.8cm}
+\sum_{gg\textrm{ states}}  \left(\Psi_{g_1 g_2}^{g_0}\right)^{\dagger} \;\;  (T^{b_f})_{b_2\, b_1}\;\;
\langle 0|\, a(\underline{p_2},\lambda_2,b_2)\, a(\underline{p_1},\lambda_1,b_1)
+\;\cdots
\bigg]. \nonumber
\end{eqnarray}
%%%%%%%%%%%%%%%%%%%%%%%%%%%%%%%%%%%%%%%%%%%%%%%%%
The different terms in this expression have the following interpretation:
\begin{itemize}
  \item First term: trivial contribution with the gluon directly emerging out of the target at $x^+=0$.
  \item Second term: contribution of heavy quark-antiquark pair merging to a gluon in the final state, see Fig. \ref{Fig:qqbar2g_FS}.
  \item Third term: contribution of gluon pair merging to one single gluon in the final state.
  \item Other terms: either they are of higher order in $g$, or they will not contribute to the  amplitudes in which we are interested.
\end{itemize}

Taking the conjugate of the wave function from Eq. \eqref{qqbar_inside_g_ISWF_mom}, and using the notations from Fig. \ref{Fig:qqbar2g_FS}, one finds
%%%%%%%%%%%%%%%%%%%%%%%%%%%%%%%%%%%%%%%%%%%%%%%%%
\begin{eqnarray}
 \left(\Psi_{q_1 \bar{q}_2}^{g_0}\right)^{\dagger}
 &=&
 \frac{ (2\pi)^{D-1} \delta^{(D-1)}(\underline{p_1} \!+\!\underline{p_2} \!-\!\underline{p_f})}
 {\bigg[\left(\p_1 \!-\! \frac{p_1^+}{p_f^+}\, \p_f\right)^2 +m^2\bigg]}
 \,  (\mu)^{2-\frac{D}{2}}\, g\,%(t^{b_f})_{\beta_2\, \beta_1}
 \nonumber\\
& & \times \Bigg\{ \left(\p_1^i \!-\! \frac{p_1^+}{p_f^+}\, \p_f^i\right) \varepsilon^{j\, *}_{\lambda_f}\; \overline{v_G}(p_2^+,h_2)\, \gamma^+ \bigg[
\frac{(p_f^+ \!-\! 2 p_1^+)}{p_f^+}\, \delta^{ij} - i\, \sigma^{ij}
 \bigg] u_G(p_1^+,h_1)\nonumber\\
& & \hspace{2cm} - m\, \varepsilon^{j\, *}_{\lambda_f}\; \overline{v_G}(p_2^+,h_2)\, \gamma^+ \gamma^j\, u_G(p_1^+,h_1)
 \Bigg\}
 \, .\label{qqbar_inside_g_FSWF_mom}
\end{eqnarray}
%%%%%%%%%%%%%%%%%%%%%%%%%%%%%%%%%%%%%%%%%%%%%%%%%
%%%%%%%%%%%%%%%%%%%%%%%%%%%%%%%%%%%%%%%%%%%%%%%%%

%%%%%%%%%%%%%%%%%%%%%%%%%%%%%%%%%%%%%%%%%%%%%%%%%%%%%%%%%%%%%%%%%%%%%%%%%%%%%%%%%%%%%%%%%%%%%%%%%%%%%%%
%%%%%%%%%%%%%%%%%%%%%%%%%%%%%%%%%%%%%%%%%%%%%%%%%%%%%%%%%%%%%%%%%%%%%%%%%%%%%%%%%%%%%%%%%%%%%%%%%%%%%%%
%%%%%%%%%%%%%%%%%%%%%%%%%%%%%%%%%%%%%%%%%%%%%%%%%%%%%%%%%%%%%%%%%%%%%%%%%%%%%%%%%%%%%%%%%%%%%%%%%%%%%%%
%%%%%%%%%%%%%%%%%%%%%%%%%%%%%%%%%%%%%%%%%%%%%%%%%%%%%%%%%%%%%%%%%%%%%%%%%%%%%%%%%%%%%%%%%%%%%%%%%%%%%%%

\subsubsection{Mixed space}

In the mixed-space representation, the one-gluon final state \eqref{phys_gFS_decomp_mom} rewrites
%%%%%%%%%%%%%%%%%%%%%%%%%%%%%%%%%%%%%%%%%%%%%%%%%
\begin{eqnarray}
\hspace{-0.5cm}
\langle g_{\textrm{phys}}(\underline{p_f},\lambda_f,b_f)|
&=& \sqrt{Z_A} \bigg[\int d^{D-2}\x_0\;\; e^{-i\p_f \cdot \x_0}\;\; \langle 0|\,
 a(p_f^+,\x_0,\lambda_f,b_f) \nonumber\\
& & \hspace{-2cm} +\widetilde{\sum_{q \bar{q}\textrm{ states}}}  \left(\widetilde{\Psi}_{q_1 \bar{q_2}}^{g_f}\right)^{\dagger}\;\;
(t^{b_f})_{\beta_2\, \beta_1}\; \;
\langle 0|\, d(p_2^+,\x_2,h_2,\beta_2)\,
b(p_1^+,\x_1,h_1,\beta_1) \nonumber\\
& & \hspace{-2cm} +\widetilde{\sum_{gg\textrm{ states}}}  \left(\widetilde{\Psi}_{g_1 g_2}^{g_f}\right)^{\dagger}\;\;  (T^{b_f})_{b_2\, b_1}\;\; \langle 0| \,
a(p_2^+,\x_2,\lambda_2,b_2)\,
a(p_1^+,\x_1,\lambda_1,b_1)
+\;\cdots
\bigg]\, ,
\label{phys_gluonFS_decomp_mix}
\end{eqnarray}
%%%%%%%%%%%%%%%%%%%%%%%%%%%%%%%%%%%%%%%%%%%%%%%%%
where
%%%%%%%%%%%%%%%%%%%%%%%%%%%%%%%%%%%%%%%%%%%%%%%%%
\begin{eqnarray}
 \left(\widetilde{\Psi}_{q_1 \bar{q}_2}^{g_f}\right)^{\dagger}
 &=& 2\pi \delta(p_1^+ \!+\!p_2^+ \!-\!p_f^+)\; e^{-i\frac{\p_f}{p^+_f}\cdot (p^+_1\, \x_1 +p^+_2 \, \x_2)}\;\;
  (\mu)^{2-\frac{D}{2}}\, g\,%(t^{b_f})_{\beta_2\, \beta_1}
 \nonumber\\
 & & \times \Bigg\{ \varepsilon^{j\, *}_{\lambda_f}\; \overline{v_G}(p_2^+,h_2)\, \gamma^+ \bigg[ \frac{(p_f^+ \!-\! 2 p_1^+)}{p_f^+}\, \delta^{ij} - i\, \sigma^{ij}
 \bigg] u_G(p_1^+,h_1)\;\;
 \left[{\cal B}^i_V(\x_{12},m)\right]^{*}\nonumber\\
 & & \hspace{2cm} -m\; \varepsilon^{j\, *}_{\lambda_f}\; \overline{v_G}(p_2^+,h_2)\, \gamma^+ \gamma^j\, u_G(p_1^+,h_1)\;\;
  \left[{\cal B}_S(\x_{12},m)\right]^{*}
 \Bigg\}\, .\label{qqbar_to_g_FS_WF}
\end{eqnarray}
%%%%%%%%%%%%%%%%%%%%%%%%%%%%%%%%%%%%%%%%%%%%%%%%%

%%%%%%%%%%%%%%%%%%%%%%%%%%%%%%%%%%%%%%%%%%%%%%%%%%%%%%%%%%%%%%%%%%%%%%%%%%%%%%%%%%%%%%%%%%%%%%%%%%%%%%%
%%%%%%%%%%%%%%%%%%%%%%%%%%%%%%%%%%%%%%%%%%%%%%%%%%%%%%%%%%%%%%%%%%%%%%%%%%%%%%%%%%%%%%%%%%%%%%%%%%%%%%%
%%%%%%%%%%%%%%%%%%%%%%%%%%%%%%%%%%%%%%%%%%%%%%%%%%%%%%%%%%%%%%%%%%%%%%%%%%%%%%%%%%%%%%%%%%%%%%%%%%%%%%%
%%%%%%%%%%%%%%%%%%%%%%%%%%%%%%%%%%%%%%%%%%%%%%%%%%%%%%%%%%%%%%%%%%%%%%%%%%%%%%%%%%%%%%%%%%%%%%%%%%%%%%%

\subsection{Heavy quark loop contribution to the gluon-to-gluon amplitude}

\subsubsection{Generic form}

In the eikonal approximation, the scattering amplitude ${\cal M}_{g\rightarrow g}$ for a gluon on the target, and the corresponding $S$-matrix element, are related by
%%%%%%%%%%%%%%%%%%%%%%%%%%%%%%%%%%%%%%%%%%%%%%%%%
\begin{eqnarray}
\langle g_{\textrm{phys}}(\underline{p_f},\lambda_f,b_f)|\; \hat{S}_E\;  |g_{\textrm{phys}}(\underline{k_0},\lambda_0,a_0)\rangle
&=& (2k_0^+)(2\pi)\delta(p_f^+\!-\!k_0^+)\; i\, {\cal M}_{g\rightarrow g}\ ,
\label{g_to_g_ampl_def}
\end{eqnarray}
%%%%%%%%%%%%%%%%%%%%%%%%%%%%%%%%%%%%%%%%%%%%%%%%%
with the eikonal scattering operator $\hat{S}_E$ introduced in subsection \ref{sec:g2qqbar_ampl}. Calculating the left-hand side using the expansions \eqref{phys_gluon_decomp_mix} and \eqref{phys_gluonFS_decomp_mix}, one finds
%%%%%%%%%%%%%%%%%%%%%%%%%%%%%%%%%%%%%%%%%%%%%%%%%
\begin{eqnarray}
& &\hspace{-1cm} \langle g_{\textrm{phys}}(\underline{p_f},\lambda_f,b_f)|\; \hat{S}_E\;  |g_{\textrm{phys}}(\underline{k_0},\lambda_0,a_0)\rangle
= Z_A\; \nonumber\\&&
\hspace{-0.5cm}
\times
\Bigg\{ (2k_0^+)(2\pi)\delta(p_f^+\!-\!k_0^+)\; \delta_{\lambda_f, \lambda_0}
%\nonumber\\
%&&\times
\int d^{D-2}\x_0\;\; e^{-i(\p_f-\k_0) \cdot \x_0}\;\; U_A(\x_0)_{b_f a_0}
\nonumber\\
&& \vspace{-1cm}
+ \widetilde{\sum_{q \bar{q}\textrm{ states}}}
\left(\widetilde{\Psi}_{q_1 \bar{q}_2}^{g_f} \right)^{\dagger}\; (t^{b_f})_{\beta_2\, \beta_1}\;
 U_F(\x_1)_{\beta_1 \alpha_1}\; \Big[U_F^{\dagger}(\x_2)\Big]_{\alpha_2 \beta_2}\;   (t^{a_0})_{\alpha_1\, \alpha_2}\;
 \widetilde{\Psi}_{q_1 \bar{q}_2}^{g_0}\label{g_to_g_ampl_1}
 \\
&& \vspace{-1cm}
+\widetilde{\sum_{gg\textrm{ states}}}
 \left( \widetilde{\Psi}_{g_1 g_2}^{g_f}\right)^{\dagger}\;
 (T^{b_f})_{b_2\, b_1}\;
U_A(\x_1)_{b_1 a_1}\;  U_A(\x_2)_{b_2 a_2}\;
\;  (T^{a_0})_{a_1\, a_2}\;
\widetilde{\Psi}_{g_1 g_2}^{g_0}
+{\cal O}\left(g^4\right)\!\!
\Bigg\}\, .
\nonumber
\end{eqnarray}
%%%%%%%%%%%%%%%%%%%%%%%%%%%%%%%%%%%%%%%%%%%%%%%%%
Isolating the leading-order contribution
%%%%%%%%%%%%%%%%%%%%%%%%%%%%%%%%%%%%%%%%%%%%%%%%%
\begin{eqnarray}
i\,{\cal M}^{LO}_{g\rightarrow g} = \delta_{\lambda_f, \lambda_0}\;
\int d^{D-2}\x_0\;\; e^{-i(\p_f-\k_0) \cdot \x_0}\;\; U_A(\x_0)_{b_f a_0}\, ,
\label{g_to_g_ampl_LO}
\end{eqnarray}
%%%%%%%%%%%%%%%%%%%%%%%%%%%%%%%%%%%%%%%%%%%%%%%%%
one rewrites eq. \eqref{g_to_g_ampl_1} as
%%%%%%%%%%%%%%%%%%%%%%%%%%%%%%%%%%%%%%%%%%%%%%%%%
\begin{eqnarray}
 (2k_0^+)(2\pi)\delta(p_f^+\!-\!k_0^+)\; i\, \left[{\cal M}_{g\rightarrow g}-{\cal M}^{LO}_{g\rightarrow g}\right]
&=&
-(1\!-\!Z_A)\; (2k_0^+)(2\pi)\delta(p_f^+\!-\!k_0^+)\; \delta_{\lambda_f, \lambda_0}
\nonumber\\
&&
\hspace{-1cm}
\times
\int d^{D-2}\x_0\;\; e^{-i(\p_f-\k_0) \cdot \x_0}\;\; U_A(\x_0)_{b_f a_0}
\nonumber\\
&&  \hspace{-6cm}
+ Z_A\; \Bigg\{\widetilde{\sum_{q \bar{q}\textrm{ states}}}
\left(\widetilde{\Psi}_{q_1 \bar{q}_2}^{g_f} \right)^{\dagger}\;
 \textrm{Tr}\big[t^{b_f}\; U_F(\x_1)\; t^{a_0} \; U_F^{\dagger}(\x_2)\big]\;
 \widetilde{\Psi}_{q_1 \bar{q}_2}^{g_0}
 \nonumber\\
&&  \hspace{-5cm}
+\widetilde{\sum_{gg\textrm{ states}}}
 \left( \widetilde{\Psi}_{g_1 g_2}^{g_f}\right)^{\dagger}\;
 \textrm{Tr}\big[T^{b_f}\;  U_A(\x_1)\; T^{a_0}\;  U_A^{\dagger}(\x_2) \big]\;
\widetilde{\Psi}_{g_1 g_2}^{g_0}
+{\cal O}\left(g^4\right)
\Bigg\}\, .
\label{g_to_g_ampl_2}
\end{eqnarray}
%%%%%%%%%%%%%%%%%%%%%%%%%%%%%%%%%%%%%%%%%%%%%%%%%

%%%%%%%%%%%%%%%%%%%%%%%%%%%%%%%%%%%%%%%%%%%%%%%%%%%%%%%%%%%%%%%%%%%%%%%%%%%%%%%%%%%%%%%%%%%%%%%%%%%%%%%
%%%%%%%%%%%%%%%%%%%%%%%%%%%%%%%%%%%%%%%%%%%%%%%%%%%%%%%%%%%%%%%%%%%%%%%%%%%%%%%%%%%%%%%%%%%%%%%%%%%%%%%
%%%%%%%%%%%%%%%%%%%%%%%%%%%%%%%%%%%%%%%%%%%%%%%%%%%%%%%%%%%%%%%%%%%%%%%%%%%%%%%%%%%%%%%%%%%%%%%%%%%%%%%
%%%%%%%%%%%%%%%%%%%%%%%%%%%%%%%%%%%%%%%%%%%%%%%%%%%%%%%%%%%%%%%%%%%%%%%%%%%%%%%%%%%%%%%%%%%%%%%%%%%%%%%

\subsubsection{Heavy quark contribution to the gluon wave function renormalization}

%%%%%%%%%%%%%%%%%%%%%%%%%%%%%%%%%%%%%%%%%%%%%%%%%%%%%%%%%%%%%%%%%%%%%%%%%%%%%%%%%%%%%%%%%%%%%%%%%%%%%%%
%%%%%%%%%%%%%%%%%%%%%%%%%%%%%%%%%%%%%%%%%%%%%%%%%%%%%%%%%%%%%%%%%%%%%%%%%%%%%%%%%%%%%%%%%%%%%%%%%%%%%%%

The gluon renormalization constant $Z_A$, appearing in the Fock state decompositions \eqref{phys_gluon_decomp_mom}, \eqref{phys_gluon_decomp_mix}, \eqref{phys_gFS_decomp_mom} and \eqref{phys_gluonFS_decomp_mix}, is determined by imposing the following normalization to the one-gluon asymptotic state:
%%%%%%%%%%%%%%%%%%%%%%%%%%%%%%%%%%%%%%%%%%%%%%%%%
\begin{eqnarray}
\langle g(\underline{p_f},\lambda_f, b_f)_{\textrm{phys}}| g(\underline{k_0},\lambda_0,a_0)_{\textrm{phys}}\rangle
&=& (2k_0^+)(2\pi)^{D\!-\!1} \delta^{(D\!-\!1)}(\underline{p_f}\!-\!\underline{k_0})\; \delta_{\lambda_f,\lambda_0}\; \delta_{b_f,a_0}\, .
\label{1_gluon_normalization}
\end{eqnarray}
%%%%%%%%%%%%%%%%%%%%%%%%%%%%%%%%%%%%%%%%%%%%%%%%%
Usually, the calculation of such renormalization constant is done in momentum space, inserting the expansions \eqref{phys_gluon_decomp_mom} and \eqref{phys_gFS_decomp_mom} into the normalization relation \eqref{1_gluon_normalization}. But we will need a mixed space relation for $Z_A$, for coherence with the rest of the calculation, so that the Fock state expansions \eqref{phys_gluon_decomp_mix} and \eqref{phys_gluonFS_decomp_mix} are used instead, and one obtains
%%%%%%%%%%%%%%%%%%%%%%%%%%%%%%%%%%%%%%%%%%%%%%%%%
\begin{eqnarray}
(2k_0^+)(2\pi)^{D\!-\!1} \delta^{(D\!-\!1)}(\underline{p_f}\!-\!\underline{k_0})\, \delta_{\lambda_f,\lambda_0}\, \delta_{b_f,a_0} \frac{(1\!-\!Z_A)}{Z_A}
\!&=&\!\!
\widetilde{\sum_{q \bar{q}\textrm{ states}}}
\left(\widetilde{\Psi}_{q_1 \bar{q}_2}^{g_f} \right)^{\dagger}\!
(t^{b_f})_{\alpha_2\, \alpha_1}\;   (t^{a_0})_{\alpha_1\, \alpha_2} \;
  \widetilde{\Psi}_{q_1 \bar{q}_2}^{g_0}  \nonumber\\
&&
\hspace{-4cm}
+\widetilde{\sum_{gg\textrm{ states}}}
\left( \widetilde{\Psi}_{g_1 g_2}^{g_f}\right)^{\dagger}\;
(T^{b_f})_{a_2\, a_1}\;\;  (T^{a_0})_{a_1\, a_2}\;
\widetilde{\Psi}_{g_1 g_2}^{g_0}
+{\cal O}\left(g^4\right)
\, ,
\label{gluon_WF_renorm_1}
\end{eqnarray}
%%%%%%%%%%%%%%%%%%%%%%%%%%%%%%%%%%%%%%%%%%%%%%%%%
thanks to the commutation relations \eqref{commute_a_adag_mix}, \eqref{anticommute_b_bdag_mix} and \eqref{anticommute_d_ddag_mix}.
Taking care of the color algebra, it results in
%%%%%%%%%%%%%%%%%%%%%%%%%%%%%%%%%%%%%%%%%%%%%%%%%
\begin{eqnarray}
(2k_0^+)(2\pi)^{D\!-\!1} \delta^{(D\!-\!1)}(\underline{p_f}\!-\!\underline{k_0})\, \delta_{\lambda_f,\lambda_0}\, \frac{(1\!-\!Z_A)}{Z_A}
&=& T_F\;
\widetilde{\sum_{q \bar{q}\textrm{ states}}}
\left(\widetilde{\Psi}_{q_1 \bar{q}_2}^{g_f} \right)^{\dagger}\;
  \widetilde{\Psi}_{q_1 \bar{q}_2}^{g_0}  \nonumber\\
&&
\hspace{-2cm}
+\: C_A\; \widetilde{\sum_{gg\textrm{ states}}}
\left( \widetilde{\Psi}_{g_1 g_2}^{g_f}\right)^{\dagger}\;
\widetilde{\Psi}_{g_1 g_2}^{g_0}
+{\cal O}\left(g^4\right)
\, .
\label{gluon_WF_renorm}
\end{eqnarray}
%%%%%%%%%%%%%%%%%%%%%%%%%%%%%%%%%%%%%%%%%%%%%%%%%
In that expression, the first term correspond to the quark loop contribution, and the second one to the gluon loop contribution. Both are of order $g^2$, obviously.
%In this work, we are only interested in the heavy quark loop contribution.
%An explicit expression for $Z_A$ won't be needed : instead, the equation \eqref{gluon_WF_renorm} will be used to substitute $Z_A$.

%%%%%%%%%%%%%%%%%%%%%%%%%%%%%%%%%%%%%%%%%%%%%%%%%%%%%%%%%%%%%%%%%%%%%%%%%%%%%%%%%%%%%%%%%%%%%%%%%%%%%%%
%%%%%%%%%%%%%%%%%%%%%%%%%%%%%%%%%%%%%%%%%%%%%%%%%%%%%%%%%%%%%%%%%%%%%%%%%%%%%%%%%%%%%%%%%%%%%%%%%%%%%%%

\subsubsection{Initial state/final state wave function overlap for a heavy quark loop}

The next step is to evaluate the quantity
%%%%%%%%%%%%%%%%%%%%%%%%%%%%%%%%%%%%%%%%%%%%%%%%%
\begin{eqnarray}
\widetilde{\sum_{q \bar{q}\textrm{ states}}}
\left(\widetilde{\Psi}_{q_1 \bar{q}_2}^{g_f} \right)^{\dagger}\;\;
\widetilde{\Psi}_{q_1 \bar{q}_2}^{g_0}\;\;  F(\x_1,\x_2)
\label{IS_FS_overlap_def}
\end{eqnarray}
%%%%%%%%%%%%%%%%%%%%%%%%%%%%%%%%%%%%%%%%%%%%%%%%%
for a generic function $F(\x_1,\x_2)$. Indeed, for $F(\x_1,\x_2)\equiv 1$, it gives the quark loop contribution to the gluon wave function renormalization \eqref{gluon_WF_renorm}, and for
\begin{equation}
F(\x_1,\x_2)\equiv \textrm{Tr}\big[t^{b_f}\, U_F(\x_1)\, t^{a_0}\, U_F^{\dagger}(\x_2)\big] \; ,
\end{equation}
it gives the resolved quark loop contribution to the scattering amplitude \eqref{g_to_g_ampl_2}.
Thanks to the expressions \eqref{g_to_qqbar_IS_WF} and \eqref{qqbar_to_g_FS_WF}, one finds
%%%%%%%%%%%%%%%%%%%%%%%%%%%%%%%%%%%%%%%%%%%%%%%%%
\begin{eqnarray}
&& \widetilde{\sum_{q \bar{q}\textrm{ states}}}
\left(\widetilde{\Psi}_{q_1 \bar{q}_2}^{g_f} \right)^{\dagger}\!
\widetilde{\Psi}_{q_1 \bar{q}_2}^{g_0}\;\;  F(\x_1,\x_2)
=
\sum_{h_1,\, h_2} \int_{0}^{+\infty}\!\!\!\! \frac{dk_1^+}{(2\pi) 2k_1^+}
 \int_{0}^{+\infty}\!\!\!\! \frac{dk_2^+}{(2\pi) 2k_2^+}
 \int \!d^{D-2} \x_1\! \int \!d^{D-2} \x_2\;\;
\nonumber\\
&& \times
F(\x_1,\x_2)
 (2\pi)\delta(k_1^+\!+\!k_2^+\!-\!k_0^+)\;  (2\pi)\delta(k_1^+\!+\!k_2^+\!-\!p_f^+)\;
e^{-i \left(\frac{\p_f}{p^+_f}-\frac{\k_0}{k^+_0}\right)\cdot(k_1^+ \x_1 + k_2^+ \x_2)}\;
(\mu^2)^{2-\frac{D}{2}}\, g^2\,
\nonumber\\
&& \times
 \varepsilon^{j'\, *}_{\lambda_f}
  \varepsilon^{j}_{\lambda_0}\; \overline{v_G}(k_2^+,h_2) \gamma^+
  \Bigg\{ {\cal B}^{i'\, *}_V(\x_{12},m)\;
  \bigg[\frac{(p_f^+ \!-\! 2 k_1^+)}{p_f^+}\, \delta^{i'j'} - i\, \sigma^{i'j'}\bigg]
  - {\cal B}^{*}_S(\x_{12},m)\; m\;  \gamma^{j'}\Bigg\}\;
\nonumber\\
 &&% \hspace{1cm}
 \times \,
  u_G(k_1^+,h_1)
  \overline{u_G}(k_1^+,h_1)\, \gamma^+ \Bigg\{ {\cal B}^{i}_V(\x_{12},m)\;
  \bigg[\frac{(k_0^+ \!-\! 2 k_1^+)}{k_0^+}\, \delta^{ij} + i\, \sigma^{ij}\bigg]
  + {\cal B}_S(\x_{12},m)\; m\;  \gamma^{j}\Bigg\}\;\,
\nonumber\\
&&\times \;  v_G(k_2^+,h_2)\\
&& =(2\pi)\delta(k_0^+\!-\!p_f^+) \int d^{D-2} \x_1\; \int d^{D-2} \x_2\;\; F(\x_1,\x_2)\;
(\mu^2)^{2-\frac{D}{2}}\, g^2\, \int_{0}^{k_0^+}\! \frac{dk_1^+}{(2\pi)}\;
\nonumber\\
&&  \times\;
e^{-i \left(\frac{\p_f-\k_0}{k^+_0}\right)\cdot(k_1^+ \x_{12} + k_0^+ \x_2)}
 \varepsilon^{j'\, *}_{\lambda_f}
  \varepsilon^{j}_{\lambda_0}\;  \Bigg\{ {\cal B}^{i'\, *}_V(\x_{12},m)\; {\cal B}^{i}_V(\x_{12},m)\nonumber\\
  &&\hspace{2cm}\times\;
   \textrm{Tr}\bigg[{{\cal P}_{G}}\, \bigg(\frac{(k_0^+ \!-\! 2 k_1^+)}{k_0^+}\, \delta^{i'j'} - i\, \sigma^{i'j'}\bigg)\, \bigg(\frac{(k_0^+ \!-\! 2 k_1^+)}{k_0^+}\, \delta^{ij} + i\, \sigma^{ij}\bigg) \bigg]
\nonumber\\
&& \hspace{2.5cm}
- \left|{\cal B}_S(\x_{12},m)\right|^2\; m^2\;
              \textrm{Tr}\bigg[{{\cal P}_{G}}\, \gamma^{j'}\, \gamma^{j} \bigg]
            \Bigg\}
\\
&& =(2k_0^+)(2\pi)\delta(k_0^+\!-\!p_f^+) \int d^{D-2} \x_1\; \int d^{D-2} \x_2\;\; F(\x_1,\x_2)\;
(\mu^2)^{2-\frac{D}{2}}\, g^2\, \int_{0}^{k_0^+}\! \frac{dk_1^+}{(2\pi)(2k_0^+)}\;
\nonumber\\
&&  \times\;
e^{-i (\p_f-\k_0)\cdot\left(\x_2 +\frac{k_1^+}{k^+_0} \x_{12}  \right)}
  \varepsilon^{j'\, *}_{\lambda_f}
  \varepsilon^{j}_{\lambda_0}\;  \Bigg\{2\, {\cal B}^{i'\, *}_V(\x_{12},m)\; {\cal B}^{i}_V(\x_{12},m)\;
  \nonumber\\
  &&
  \hspace{1cm}
  \times
  \bigg[ \left(\frac{k_0^+ \!-\! 2 k_1^+}{k_0^+}\right)^2\, \delta^{i'j'}\,  \delta^{ij}
  -\delta^{i'j'}\,  \delta^{ij} +  \delta^{i'i}\,  \delta^{j'j}
  \bigg]
%\nonumber\\
%&&
+2\, m^2\; \delta^{j'j}\;  \left|{\cal B}_S(\x_{12},m)\right|^2
            \Bigg\}
\, ,
\label{IS_FS_overlap_1}
\end{eqnarray}
%%%%%%%%%%%%%%%%%%%%%%%%%%%%%%
using, in order to simplify the calculation of the Dirac trace, the fact that ${\cal B}^{i'\, *}_V(\x_{12},m)\; {\cal B}^{i}_V(\x_{12},m)$ is invariant under the exchange of $i$ and $i'$.

%%%%%%%%%%%%%%%%%%%%%%%%%%%%%%%%%%%%%%%%%%%%%%%%%%%%%%%%%%%%%%%%%%%%%%%%%%%%%%%%%%%%%%%%%%%%%%%%%%%%%%%
%%%%%%%%%%%%%%%%%%%%%%%%%%%%%%%%%%%%%%%%%%%%%%%%%%%%%%%%%%%%%%%%%%%%%%%%%%%%%%%%%%%%%%%%%%%%%%%%%%%%%%%
%%%%%%%%%%%%%%%%%%%%%%%%%%%%%%%%%%%%%%%%%%%%%%%%%%%%%%%%%%%%%%%%%%%%%%%%%%%%%%%%%%%%%%%%%%%%%%%%%%%%%%%
%%%%%%%%%%%%%%%%%%%%%%%%%%%%%%%%%%%%%%%%%%%%%%%%%%%%%%%%%%%%%%%%%%%%%%%%%%%%%%%%%%%%%%%%%%%%%%%%%%%%%%%

\subsubsection{Back to the heavy quark contribution to the gluon wave function renormalization}

For the case $F(\x_1,\x_2)\equiv 1$ one can perform the integration over $\x_2$ in \eqref{IS_FS_overlap_1}, while keeping $\x_{12}$ as independent integration variable. One gets
%%%%%%%%%%%%%%%%%%%%%%%%%%%%%%%%%%%%%%%%%%%%%%%%%
\begin{eqnarray}
&&\hspace{-1.5cm} \widetilde{\sum_{q \bar{q}\textrm{ states}}}
\left(\widetilde{\Psi}_{q_1 \bar{q}_2}^{g_f} \right)^{\dagger}\;\;
\widetilde{\Psi}_{q_1 \bar{q}_2}^{g_0}\;\;
 =(2k_0^+)(2\pi)^{D\!-\!1} \delta^{(D\!-\!1)}(\underline{p_f}\!-\!\underline{k_0})\; \int d^{D-2} \x_{12}\;\;
(\mu^2)^{2-\frac{D}{2}}\,
\nonumber\\
&&\times\;
g^2\, \int_{0}^{k_0^+}\! \frac{dk_1^+}{(2\pi)(2k_0^+)}\;
%\nonumber\\
%&&  \times\;\;
 \varepsilon^{j'\, *}_{\lambda_f}
  \varepsilon^{j}_{\lambda_0}\;  \Bigg\{2\, {\cal B}^{i'\, *}_V(\x_{12},m)\; {\cal B}^{i}_V(\x_{12},m)
  \nonumber\\
  &&\times
  \bigg[ \left(\frac{k_0^+ \!-\! 2 k_1^+}{k_0^+}\right)^2\, \delta^{i'j'}\,  \delta^{ij}
  -\delta^{i'j'}\,  \delta^{ij} +  \delta^{i'i}\,  \delta^{j'j}
  \bigg]
%\nonumber\\
%&&
+2\, m^2\; \delta^{j'j}\;  \left|{\cal B}_S(\x_{12},m)\right|^2
            \Bigg\}
\, .
\label{IS_FS_overlap_2}
\end{eqnarray}
%%%%%%%%%%%%%%%%%%%%%%%%%%%%%%%%%%%%%%%%%%%%%%%%%
Hence, the heavy quark loop contribution to the wave function renormalization obtained from eq. \eqref{gluon_WF_renorm} is
%%%%%%%%%%%%%%%%%%%%%%%%%%%%%%%%%%%%%%%%%%%%%%%%%
\begin{eqnarray}
&&
\hspace{-1.5cm}
 \delta_{\lambda_f,\lambda_0}\, \frac{(1\!-\!Z_A)}{Z_A}\bigg|_{\textrm{quark loop}}
 =(\mu^2)^{2-\frac{D}{2}}\, g^2\, T_F\; \int d^{D-2} \x_{12}\;\;
 \int_{0}^{k_0^+}\! \frac{dk_1^+}{(2\pi)(2k_0^+)}\;
\varepsilon^{j'\, *}_{\lambda_f}
  \varepsilon^{j}_{\lambda_0}
\nonumber\\
&&  \times\; \Bigg\{2\, {\cal B}^{i'\, *}_V(\x_{12},m)\; {\cal B}^{i}_V(\x_{12},m)\;
  \bigg[ \left(\frac{k_0^+ \!-\! 2 k_1^+}{k_0^+}\right)^2\, \delta^{i'j'}\,  \delta^{ij}
  -\delta^{i'j'}\,  \delta^{ij} +  \delta^{i'i}\,  \delta^{j'j}
  \bigg]
\nonumber\\
&&
\hspace{0.8cm}
+2\, m^2\; \delta^{j'j}\;  \left|{\cal B}_S(\x_{12},m)\right|^2
            \Bigg\}
\, .
\label{gluon_WF_renorm_quark_loop}
\end{eqnarray}
%%%%%%%%%%%%%%%%%%%%%%%%%%%%%%%%%%%%%%%%%%%%%%%%%

%%%%%%%%%%%%%%%%%%%%%%%%%%%%%%%%%%%%%%%%%%%%%%%%%%%%%%%%%%%%%%%%%%%%%%%%%%%%%%%%%%%%%%%%%%%%%%%%%%%%%%%
%%%%%%%%%%%%%%%%%%%%%%%%%%%%%%%%%%%%%%%%%%%%%%%%%%%%%%%%%%%%%%%%%%%%%%%%%%%%%%%%%%%%%%%%%%%%%%%%%%%%%%%
%%%%%%%%%%%%%%%%%%%%%%%%%%%%%%%%%%%%%%%%%%%%%%%%%%%%%%%%%%%%%%%%%%%%%%%%%%%%%%%%%%%%%%%%%%%%%%%%%%%%%%%
%%%%%%%%%%%%%%%%%%%%%%%%%%%%%%%%%%%%%%%%%%%%%%%%%%%%%%%%%%%%%%%%%%%%%%%%%%%%%%%%%%%%%%%%%%%%%%%%%%%%%%%

\subsubsection{Explicit expression for the heavy quark loop correction}

Inserting the expressions \eqref{IS_FS_overlap_1} and \eqref{gluon_WF_renorm_quark_loop} into the general expression \eqref{g_to_g_ampl_2} for the gluon to gluon scattering amplitude, and dropping the gluon loop contributions (and higher orders), one finds\footnote{In order to get a more compact expression, the integration variable $\x_0$ has been relabelled $\x_1$.}
%%%%%%%%%%%%%%%%%%%%%%%%%%%%%%%%%%%%%%%%%%%%%%%%%
\begin{eqnarray}
& &
\hspace{-0.5cm}
 i\, {\cal M}_{g\rightarrow g}\bigg|_{\textrm{quark loop}}
=
(\mu^2)^{2-\frac{D}{2}}\, g^2\, \int_{0}^{k_0^+}\! \frac{dk_1^+}{(2\pi)(2k_0^+)}\;
 \int d^{D-2} \x_1\; \int d^{D-2} \x_2\;\;\; \varepsilon^{j'\, *}_{\lambda_f}
  \varepsilon^{j}_{\lambda_0}
\nonumber\\
&&  \times\;
\Bigg\{2\, {\cal B}^{i'\, *}_V(\x_{12},m)\; {\cal B}^{i}_V(\x_{12},m)\;
  \bigg[ \left(\frac{k_0^+ \!-\! 2 k_1^+}{k_0^+}\right)^2\, \delta^{i'j'}\,  \delta^{ij}
  -\delta^{i'j'}\,  \delta^{ij} +  \delta^{i'i}\,  \delta^{j'j}
  \bigg]
\nonumber\\
&&\hspace{1cm}
+2\, m^2\; \delta^{j'j}\;  \left|{\cal B}_S(\x_{12},m)\right|^2
            \Bigg\}\label{g_to_g_ampl_quark_loop}
\\
&&
\hspace{-0.5cm}
\times\;
\Bigg\{
\textrm{Tr}\big[t^{b_f}\; U_F(\x_1)\; t^{a_0} \; U_F^{\dagger}(\x_2)\big]\;
e^{-i (\p_f-\k_0)\cdot\left(\x_2 +\frac{k_1^+}{k^+_0} \x_{12}  \right)}
- T_F\;\; U_A(\x_1)_{b_f a_0} \; e^{-i(\p_f-\k_0) \cdot \x_1}
\Bigg\}
\, .\nonumber
\end{eqnarray}
%%%%%%%%%%%%%%%%%%%%%%%%%%%%%%%%%%%%%%%%%%%%%%%%%
Note that the two terms in the last line cancel one each other when $\x_1$ and $\x_2$ coincide. Hence, there is a cancelation of UV divergences between the resolved quark loop graph and the quark loop contributions to the gluon wave function renormalization, leaving the UV finite result \eqref{g_to_g_ampl_quark_loop}. This is to be expected since $Z_A$ is determined by unitarity, without introducing a counterterm in the lagrangian (or hamiltonian).

%%%%%%%%%%%%%%%%%%%%%%%%%%%%%%%%%%%%%%%%%%%%%%%%%%%%%%%%%%%%%%%%%%%%%%%%%%%%%%%%%%%%%%%%%%%%%%%%%%%%%%%
%%%%%%%%%%%%%%%%%%%%%%%%%%%%%%%%%%%%%%%%%%%%%%%%%%%%%%%%%%%%%%%%%%%%%%%%%%%%%%%%%%%%%%%%%%%%%%%%%%%%%%%
%%%%%%%%%%%%%%%%%%%%%%%%%%%%%%%%%%%%%%%%%%%%%%%%%%%%%%%%%%%%%%%%%%%%%%%%%%%%%%%%%%%%%%%%%%%%%%%%%%%%%%%
%%%%%%%%%%%%%%%%%%%%%%%%%%%%%%%%%%%%%%%%%%%%%%%%%%%%%%%%%%%%%%%%%%%%%%%%%%%%%%%%%%%%%%%%%%%%%%%%%%%%%%%

\section{Heavy quark contribution to the NLO correction to the single inclusive hadron production cross section}
\label{sec5}

%%%%%%%%%%%%%%%%%%%%%%%%%%%%%%%%%%%%%%%%%%%%%%%%%%%%%%%%%%%%%%%%%%%%%%%%%%%%%%%%%%%%%%%%%%%%%%%%%%%%%%%
%%%%%%%%%%%%%%%%%%%%%%%%%%%%%%%%%%%%%%%%%%%%%%%%%%%%%%%%%%%%%%%%%%%%%%%%%%%%%%%%%%%%%%%%%%%%%%%%%%%%%%%

\subsection{Heavy quark loop contribution to the partonic cross section}

Here, we are interested in the heavy quark loop contribution to the partonic cross section for the process $g+A\rightarrow g+X$. It is obtained from the overlap of the  heavy quark loop contribution \eqref{g_to_g_ampl_quark_loop} to the $g+A\rightarrow g+X$ amplitude with the LO contribution \eqref{g_to_g_ampl_LO}, as
%%%%%%%%%%%%%%%%%%%%%%%%%%%%%%%%%%%%%%%%%%%%%%%%%
\begin{eqnarray}
(2p_f^+)(2\pi)^{D\!-\!1} \frac{d\sigma^{g+A\rightarrow g+X}}{dp_f^+\, d^{D\!-\!2}\p_f}
\bigg|_{\textrm{quark loop}}
&=&  (2k^+_0)(2\pi)\delta(p_f^+\!-\! k_0^+)\; \nonumber\\
&&
\hspace{-4cm}
\times \;
\frac{1}{d_A}\sum_{a_0,\, b_f} \frac{1}{D\!-\!2}\sum_{\lambda_0,\, \lambda_f}
\bigg\{   \left(i\,{\cal M}^{LO}_{g\rightarrow g}\right)^{\dagger}\;
          i\, {\cal M}_{g\rightarrow g}\Big|_{\textrm{quark loop}}
        \;\; \; + \; c.c.
\bigg\}
\, .
\end{eqnarray}
%%%%%%%%%%%%%%%%%%%%%%%%%%%%%%%%%%%%%%%%%%%%%%%%%
A straightforward calculation gives
%%%%%%%%%%%%%%%%%%%%%%%%%%%%%%%%%%%%%%%%%%%%%%%%%
\begin{eqnarray}
&& (2p_f^+)(2\pi)^{D\!-\!1} \frac{d\sigma^{g+A\rightarrow g+X}}{dp_f^+\, d^{D\!-\!2}\p_f}
\bigg|_{\textrm{quark loop}}
=  (2k^+_0)(2\pi)\delta(p_f^+\!-\! k_0^+)\;  \alpha_s\, T_F
\int_{0}^{k_0^+}\! \frac{dk_1^+}{k_0^+}\; (\mu^2)^{2-\frac{D}{2}}\nonumber\\
&&\times \;
\int d^{D-2} \x_0\; \int d^{D-2} \x_1\;
%\nonumber\\
%&&  \times\;
\int d^{D-2} \x_2
\Bigg\{ \Bigg[\frac{4}{D\!-\!2}\bigg[\left(\frac{k_1^+}{k_0^+}\right)^2+ \left(\frac{k_0^+ \!-\!  k_1^+}{k_0^+}\right)^2 +\frac{D\!-\!4}{2}\bigg]
\nonumber\\
&&
\hspace{2cm}
\times \;
{\cal B}^{i\, *}_V(\x_{12},m)\; {\cal B}^{i}_V(\x_{12},m)\;
%\nonumber\\
%&&
+2\, m^2\;  \left|{\cal B}_S(\x_{12},m)\right|^2
            \Bigg]
\nonumber\\
&&  \times\;
\bigg[S_{120}\;
e^{-i (\p_f-\k_0)\cdot\left(\x_{20} +\frac{k_1^+}{k^+_0} \x_{12}  \right)}
-  S^A_{10}\; e^{-i(\p_f-\k_0) \cdot \x_{10}}
\bigg]
\;\; \; + \; c.c.\Bigg\}
\, .\label{part_cross_sec_qq_loop}
\end{eqnarray}
%%%%%%%%%%%%%%%%%%%%%%%%%%%%%%%%%%%%%%%%%%%%%%%%%
%The large mass limit of the heavy quark loop contribution to the partonic cross section can be written, by using eq. \eqref{BlargeM}, as
%%
%\begin{eqnarray}
%&&
%\hspace{-1cm}
%(2p_f^+)(2\pi)^{D\!-\!1} \frac{d\sigma^{g+A\rightarrow g+X}}{dp_f^+\, d^{D\!-\!2}\p_f}
%\bigg|_{\textrm{quark loop}}
%\!\!\!\simeq (2k^+_0)(2\pi)\delta(p_f^+\!-\! k_0^+)\;  \alpha_s\, T_F
%\int_{0}^{k_0^+}\! \frac{dk_1^+}{k_0^+}\; (\mu^2)^{2-\frac{D}{2}}\nonumber\\
%&&
%\hspace{-0.5cm}
%\times\frac{2}{m^2}\delta^{(D-2)}(0)\int d^{D-2}\x_0 \int d^{D-2}\b_{12}\;
%e^{-i(\p_f-\k_0)\cdot(\b_{12}-\x_0)}
%\nonumber\\
%&&
%\hspace{-0.5cm}
%\times
%\left\{
%\frac{1}{d_F\,C_F}{\rm Tr}\left[U_F(\b_{12})t^aU_F^{\dagger}(\b_{12})t^b\right]U_A(\x_0)_{ba}\right.
%\left.-\frac{1}{C_A}{\rm Tr}\left[U_A(\b_{12})U^{\dagger}_A(\x_0)\right]\right\}+c.c. \, .
%\end{eqnarray}
%%
%

%%%%%%%%%%%%%%%%%%%%%%%%%%%%%%%%%%%%%%%%%%%%%%%%%%%%%%%%%%%%%%%%%%%%%%%%%%%%%%%%%%%%%%%%%%%%%%%%%%%%%%%
%%%%%%%%%%%%%%%%%%%%%%%%%%%%%%%%%%%%%%%%%%%%%%%%%%%%%%%%%%%%%%%%%%%%%%%%%%%%%%%%%%%%%%%%%%%%%%%%%%%%%%%

\subsection{Hadron-level cross section}

The NLO corrections calculated in  refs. \cite{bowen,we} for the single inclusive hadron production considered only the contributions from gluons and massless quarks. However, the inclusion of heavy quarks leads to additional NLO contributions.

First, there is the possibility of producing a heavy quark, which then fragments for example into a pion or an unidentified hadron (depending on the precise observable that one is considering). That contribution is given by the sum of the results \eqref{hadr_cross_sec_befbef}, \eqref{hadr_cross_sec_aftaft} and \eqref{hadr_cross_sec_interf}, up to the appropriate change of fragmentation function.

Second, there is the contribution from heavy quark loops. At partonic level, this corresponds to eq. \eqref{part_cross_sec_qq_loop}. In order to transform it into a hadron level cross section, one can follow the same steps as in subsection \ref{sec:hadr_level_cross_sec_real}, in particular eq. \eqref{hadr_cross_sec_def}. One obtains
%%%%%%%%%%%%%%%%%%%%%%%%%%%%%%%%%%%%%%%%%%%%%%%%%
\begin{eqnarray}
 &&
 \hspace{-0.5cm}
 (2\pi)^{D\!-\!2}\, \frac{d\sigma^{p+A\rightarrow h+X}}{d x_F\, d^{D\!-\!2}\p_h}\bigg|_{\textrm{quark loop}}
= \int_{0}^{1} d x_B\;  g(x_B,\mu^2)  \int_{0}^{1}\frac{d\zeta}{\zeta^{D-2}}\; D_{h/g}(\zeta,\mu^2)\;\;\;
 \alpha_s\, T_F\; \nonumber\\
 &&\times\;   x_B\, \delta\left(x_B\!-\! \frac{x_F}{\zeta}\right) \; \int_{0}^{1} dz\;
\int d^{D\!-\!2} \x_0   \int d^{D\!-\!2} \x_1\;\int d^{D\!-\!2} \x_{2}  \;
 \nonumber\\
&&
%\hspace{-5cm}
\times \;
\Bigg\{\left[\frac{4}{D\!-\!2}\left[z^2+ (1\!-\!z)^2 +\frac{D\!-\!4}{2}\right]\, {\cal B}^{i\, *}_V(\x_{12},m)\; {\cal B}^{i}_V(\x_{12},m)\;
%\nonumber\\
%&&
+2\, m^2\;  \left|{\cal B}_S(\x_{12},m)\right|^2
            \right]
\nonumber\\
&& \;\;\;\; \times\;
\bigg[S_{120}\;
e^{-\frac{i}{\zeta}\, \p_h\cdot\left(\x_{20} +z\, \x_{12}  \right)}
-  S^A_{10}\; e^{-\frac{i}{\zeta}\,\p_h \cdot \x_{10}}
\bigg]
\;\; \; + \; c.c.\Bigg\}
\, ,\label{hadr_cross_sec_qq_loop}
\end{eqnarray}
%%%%%%%%%%%%%%%%%%%%%%%%%%%%%%%%%%%%%%%%%%%%%%%%%
{ with $D_{h/g}(\zeta,\mu^2)$ the fragmentation function for gluon into pion or unidentified hadron.}

{ On the other hand, one also expects a contribution from intrinsic charm or bottom in the proton. At LO, it writes the same as in the equation \eqref{hadr_cross_sec_intr_charm_LO}, but with the fragmentation function now into pion or unidentified hadron. This contribution is expected to be suppressed by the smallness of the intrinsic heavy flavor PDF $Q(x_B)$. But it is formally LO instead of NLO.
Hence, if this contribution is not negligible, one may want to include also the corresponding NLO corrections, whose calculation is however beyond the scope of the present study.}
%The large mass limit of the heavy quark loop contribution to the hadronic cross section reads
%%
%\begin{eqnarray}
%&&
% \hspace{-1cm}
% (2\pi)^{D\!-\!2}\, \frac{d\sigma^{p+A\rightarrow h+X}}{d x_F\, d^{D\!-\!2}\p_h}\bigg|_{\textrm{quark loop}}
%\simeq \int_{0}^{1} d x_B\;  g(x_B,\mu^2)  \int_{0}^{1}\frac{d\zeta}{\zeta^{D-2}}\; D_{h/q}(\zeta,\mu^2)\;\;\;
% \alpha_s\, T_F\; \nonumber\\
% &&
% \hspace{-1cm}
% \times\;   x_B\, \delta\left(x_B\!-\! \frac{x_F}{\zeta}\right)
% \frac{2}{m^2}\, \delta^{(D-2)}(0)
% \; \int_{0}^{1} dz\;
%\int d^{D\!-\!2} \x_0   \int d^{D\!-\!2} \b_{12}
%\; e^{-\frac{i}{\zeta}\p_h\cdot(\b_{12}-\x_0)}
% \nonumber\\
%&&
%\hspace{-1cm}
%\times
%\Bigg\{\!
%\left[
%\frac{1}{d_F\, C_F}{\rm Tr}\left[U_F(\b_{12})t^aU^{\dagger}_F(\b_{12})t^b\right]U_A(\x_0)_{ba}-\frac{1}{d_A}{\rm Tr}\left[U_A(\b_{12})U^{\dagger}_A(\x_0)\right]\right]+c.c. \Bigg\} \, .
%\end{eqnarray}
%%
%

%%%%%%%%%%%%%%%%%%%%%%%%%%%%%%%%%%%%%%%%%%%%%%%%%%%%%%%%%%%%%%%%%%%%%%%%%%%%%%%%%%%%%%%%%%%%%%%%%%%%%%%
%%%%%%%%%%%%%%%%%%%%%%%%%%%%%%%%%%%%%%%%%%%%%%%%%%%%%%%%%%%%%%%%%%%%%%%%%%%%%%%%%%%%%%%%%%%%%%%%%%%%%%%
%%%%%%%%%%%%%%%%%%%%%%%%%%%%%%%%%%%%%%%%%%%%%%%%%%%%%%%%%%%%%%%%%%%%%%%%%%%%%%%%%%%%%%%%%%%%%%%%%%%%%%%
%%%%%%%%%%%%%%%%%%%%%%%%%%%%%%%%%%%%%%%%%%%%%%%%%%%%%%%%%%%%%%%%%%%%%%%%%%%%%%%%%%%%%%%%%%%%%%%%%%%%%%%

\section*{Acknowledgments}
%M.L. thanks the Physics Department of the University of Connecticut for hospitality at the time when
%this project was initiated and completed; N.A. and A.K. thank the Physics Department of the
% Ben-Gurion University of the Negev, for warm hospitality during stays when parts of this work were done.
% ML thanks UConn, and NA and AK BGU, for warm hospitality during stays when parts of this work were done.
%This research  was supported by the EU FP7  grant  \#318921; the DOE grant DE-FG02-13ER41989; the BSF grant \#2012124, the  ISF grant \#87277111;  the ERC grant HotLHC ERC-2011-StG-279579, Xunta de Galicia (Programa Incite),  the Spanish Consolider-Ingenio 2010 Programme CPAN and  FEDER.
We express our gratitude to the Department of Physics of Ben-Gurion University of the Negev, for warm hospitality during stays when parts of this work were done (TA, NA and AK) and for financial support as Distinguished Scientist Visitor (NA).
 ML thanks the Physics Department of the University of Connecticut for hospitality.
This research  was supported by the People Programme (Marie Curie Actions) of the European Union's Seventh Framework Programme FP7/2007-2013/ under REA
grant agreement \#318921; the DOE grant DE-FG02-13ER41989 (AK); the BSF grant \#2012124 (ML and AK); the Kreitman Foundation (GB);  the Israeli Science Foundation grant \#87277111 (GB and ML); and the European Research Council grant HotLHC ERC-2011-StG-279579, Ministerio de Ciencia e Innovaci\'on of Spain under project FPA2014-58293-C2-1-P and Xunta de Galicia (Conseller\'{\i}a de Educaci\'on) within the Strategic Unit AGRUP2015/11 (TA and NA).

\appendix

\section{Conventions}
\label{appendix}

\subsection{Fock space and the interaction picture}

In the interaction picture the quark and gluon fields read
%%%%%%%%%%%%%%%%%%%%%%%%%%%%%%%%%%%%%%%%%%%%%%%%%
\begin{eqnarray}
\Psi_{\alpha}(x)&=&\int_{0}^{+\infty}\!\!\!\! \frac{dk^+}{(2\pi) 2k^+}
 \int \frac{d^{D-2} \k}{(2\pi)^{D-2}}
 \\
 &&\times \;
 \sum_{h=\pm \frac{1}{2}}
 \bigg[e^{-ik\cdot x}\, b(\underline{k},h,\alpha)\, u(\underline{k},h)
 +e^{+ik\cdot x}\, d^{\dagger}(\underline{k},h,\alpha)\, v(\underline{k},h)
 \bigg]\Bigg|_{k^-\equiv \frac{\k^2+m^2}{2k^+}}\ ,\nonumber
 \label{quant_free_q}
\end{eqnarray}
%%%%%%%%%%%%%%%%%%%%%%%%%%%%%%%%%%%%%%%%%%%%%%%%%
%Gluon field in the interaction picture:
%%%%%%%%%%%%%%%%%%%%%%%%%%%%%%%%%%%%%%%%%%%%%%%%%
\begin{eqnarray}
A_{a}^{\mu}(x)&=&\int_{0}^{+\infty}\!\!\!\! \frac{dk^+}{(2\pi) 2k^+}
 \int \frac{d^{D-2} \k}{(2\pi)^{D-2}} \nonumber\\
 &&\times \;
\sum_{\lambda} \bigg[e^{-ik\cdot x}\, a(\underline{k},\lambda,a)\, \epsilon^{\mu}_{\lambda}\!(\underline{k})
 +e^{+ik\cdot x}\, a^{\dagger}(\underline{k},\lambda,a)\, \epsilon^{\mu\, *}_{\lambda}\!(\underline{k})
 \bigg]\Bigg|_{k^-\equiv \frac{\k^2}{2k^+}}\ ,
  \label{quant_free_g}
\end{eqnarray}
%%%%%%%%%%%%%%%%%%%%%%%%%%%%%%%%%%%%%%%%%%%%%%%%%
respectively.

The commutation relations for creation and annihilation operators read
%%%%%%%%%%%%%%%%%%%%%%%%%%%%%%%%%%%%%%%%%%%%%%%%%
\begin{eqnarray}
\big[a(\underline{k_1},\lambda_1,a_1),a^{\dagger}(\underline{k_2},\lambda_2,a_2)\big]
&=& (2k_1^+)(2\pi)^{D\!-\!1}\delta^{(D\!-\!1)}(\underline{k_1}\!-\!\underline{k_2})\;
\delta_{\lambda_1,\lambda_2}\; \delta_{a_1,a_2\ ,}\\
\big\{b(\underline{k_1},h_1,\alpha_1),b^{\dagger}(\underline{k_2},h_2,\alpha_2)\big\}
&=& (2k_1^+)(2\pi)^{D\!-\!1}\delta^{(D\!-\!1)}(\underline{k_1}\!-\!\underline{k_2})\;
\delta_{h_1,h_2}\; \delta_{\alpha_1,\alpha_2}\ ,\\
\big\{d(\underline{k_1},h_1,\alpha_1),d^{\dagger}(\underline{k_2},h_2,\alpha_2)\big\}
&=& (2k_1^+)(2\pi)^{D\!-\!1}\delta^{(D\!-\!1)}(\underline{k_1}\!-\!\underline{k_2})\;
\delta_{h_1,h_2}\; \delta_{\alpha_1,\alpha_2}\ .
\end{eqnarray}
%%%%%%%%%%%%%%%%%%%%%%%%%%%%%%%%%%%%%%%%%%%%%%%%%

The Fourier transform from momentum space to mixed space is defined, for the gluon creation operator, as follows:
%%%%%%%%%%%%%%%%%%%%%%%%%%%%%%%%%%%%%%%%%%%%%%%%%
\begin{eqnarray}
a^{\dagger}(\underline{k},\lambda,a)=\int d^{D-2}\x\;\; e^{i\k \cdot \x}\;\;
a^{\dagger}(k^+,\x,\lambda,a)\ ,
\label{FT_a_dag}
\end{eqnarray}
%%%%%%%%%%%%%%%%%%%%%%%%%%%%%%%%%%%%%%%%%%%%%%%%%
and analogously for other quantities.

Hence, we have the mixed-space commutation relations
%%%%%%%%%%%%%%%%%%%%%%%%%%%%%%%%%%%%%%%%%%%%%%%%%
\begin{eqnarray}
\big[a(k_1^+,\x_1,\lambda_1,a_1),a^{\dagger}(k_2^+,\x_2,\lambda_2,a_2)\big]
&=& {\cal D}(k_1^+,k_2^+)\;\delta^{(D\!-\!2)}(\x_1\!-\!\x_2)\;
\delta_{\lambda_1,\lambda_2}\; \delta_{a_1,a_2}\, ,\label{commute_a_adag_mix}\\
\big\{b(k_1^+,\x_1,h_1,\alpha_1),b^{\dagger}(k_2^+,\x_2,h_2,\alpha_2)\big\}
&=& {\cal D}(k_1^+,k_2^+)\;\delta^{(D\!-\!2)}(\x_1\!-\!\x_2)\;
\delta_{h_1,h_2}\; \delta_{\alpha_1,\alpha_2}\, ,\label{anticommute_b_bdag_mix}\\
\big\{d(k_1^+,\x_1,h_1,\alpha_1),d^{\dagger}(k_2^+,\x_2,h_2,\alpha_2)\big\}
&=& {\cal D}(k_1^+,k_2^+)\;\delta^{(D\!-\!2)}(\x_1\!-\!\x_2)\;
\delta_{h_1,h_2}\; \delta_{\alpha_1,\alpha_2}\, ,\label{anticommute_d_ddag_mix}
\end{eqnarray}
%%%%%%%%%%%%%%%%%%%%%%%%%%%%%%%%%%%%%%%%%%%%%%%%%
with
\begin{equation}
{\cal D}(k_1^+,k_2^+)=(2k_1^+)(2\pi)\delta(k_1^+\!-\!k_2^+).
\end{equation}

\subsection{Spinors}

The projectors over good ($G$) and bad ($B$) components of a spinor $\Psi$ are defined
%%%%%%%%%%%%%%%%%%%%%%%%%%%%%%%%%%%%%%%%%%%%%%%%%
\begin{eqnarray}
{\cal P}_{G}\equiv \frac{\gamma^-\, \gamma^+}{2} = \frac{\gamma^0\, \gamma^+}{\sqrt{2}}\ ,
\nonumber \\
{\cal P}_{B}\equiv \frac{\gamma^+\, \gamma^-}{2} = \frac{\gamma^0\, \gamma^-}{\sqrt{2}}\ ,
\end{eqnarray}
%%%%%%%%%%%%%%%%%%%%%%%%%%%%%%%%%%%%%%%%%%%%%%%%%
so
%%%%%%%%%%%%%%%%%%%%%%%%%%%%%%%%%%%%%%%%%%%%%%%%%
\begin{equation}
\Psi_{G,B}   \equiv {\cal P}_{G,B}\; \Psi\ .
\end{equation}
%%%%%%%%%%%%%%%%%%%%%%%%%%%%%%%%%%%%%%%%%%%%%%%%%
Note that
%%%%%%%%%%%%%%%%%%%%%%%%%%%%%%%%%%%%%%%%%%%%%%%%%
\begin{equation}
\overline{\Psi} \;  {\cal P}_{B}     = \overline{\Psi_{G}},\ \
\overline{\Psi}  \; {\cal P}_{G}     = \overline{\Psi_{B}}.
\end{equation}
%%%%%%%%%%%%%%%%%%%%%%%%%%%%%%%%%%%%%%%%%%%%%%%%%

Concerning the solutions $u(\underline{k},h)$ and $v(\underline{k},h)$ of the free Dirac equation, the good and bad components are related through
%%%%%%%%%%%%%%%%%%%%%%%%%%%%%%%%%%%%%%%%%%%%%%%%%
\begin{eqnarray}
u_B(\underline{k},h) &=& \frac{\gamma^+}{2 k^+}\, \left(\k^j \gamma^j \!+\! m \right)\, u_G(k^+,h), \label{u_bad}\nonumber\\
v_B(\underline{k},h) &=& \frac{\gamma^+}{2 k^+}\, \left(\k^j \gamma^j \!-\! m \right)\, v_G(k^+,h). \label{v_bad}
\end{eqnarray}
%%%%%%%%%%%%%%%%%%%%%%%%%%%%%%%%%%%%%%%%%%%%%%%%%
In this way, the dependence on $\k$ and $m$ appears only in the bad components:
%%%%%%%%%%%%%%%%%%%%%%%%%%%%%%%%%%%%%%%%%%%%%%%%%
\begin{eqnarray}
\overline{u_B}(\underline{k},h) &=&  \overline{u_G}(k^+,h)\, \left(\k^j \gamma^j \!+\! m \right)\,  \frac{\gamma^+}{2 k^+}\ ,  \label{ubar_bad}\nonumber\\
\overline{v_B}(\underline{k},h) &=& \overline{v_G}(k^+,h)\, \left(\k^j \gamma^j \!-\! m \right)\,\frac{\gamma^+}{2 k^+}\ . \label{vbar_bad}
\end{eqnarray}
%%%%%%%%%%%%%%%%%%%%%%%%%%%%%%%%%%%%%%%%%%%%%%%%%

Finally, the completeness relations read
%%%%%%%%%%%%%%%%%%%%%%%%%%%%%%%%%%%%%%%%%%%%%%%%%
\begin{eqnarray}
\sum_{h=\pm\frac{1}{2}} u_G(k^+,h)\;  \overline{u_G}(k^+,h)\, \gamma^+
&=& \sum_{h=\pm\frac{1}{2}} v_G(k^+,h)\;  \overline{v_G}(k^+,h)\, \gamma^+
= 2k^+\; {\cal P}_{G}\ .
\end{eqnarray}
%%%%%%%%%%%%%%%%%%%%%%%%%%%%%%%%%%%%%%%%%%%%%%%%%

\subsection{Polarization vectors}

The polarization vectors in light-cone gauge $A^+=0$ are defined
%%%%%%%%%%%%%%%%%%%%%%%%%%%%%%%%%%%%%%%%%%%%%%%%%
\begin{eqnarray}
\epsilon^{+}_{\lambda}\!(\underline{k})&=& 0, \nonumber\\
\epsilon^{j}_{\lambda}\!(\underline{k})&=& \varepsilon^{j}_{\lambda}\ ,\nonumber\\
\epsilon^{-}_{\lambda}\!(\underline{k})&=& \frac{\k^j\, \varepsilon^{j}_{\lambda}}{k^+}\ ,
\label{4_to_2_polarization}
\end{eqnarray}
%%%%%%%%%%%%%%%%%%%%%%%%%%%%%%%%%%%%%%%%%%%%%%%%%
where the transverse vectors $\varepsilon_{\lambda}$ obey the relations
%%%%%%%%%%%%%%%%%%%%%%%%%%%%%%%%%%%%%%%%%%%%%%%%%
\begin{eqnarray}
\sum_{\lambda}\varepsilon^{i}_{\lambda}\, \varepsilon^{j\, *}_{\lambda}&=& -g^{ij},\nonumber\\
-g_{ij}\, \varepsilon^{i}_{\lambda_1}\, \varepsilon^{j\, *}_{\lambda_2}&=& \delta_{\lambda_1,\lambda_2}\, .
\label{Rel_polarizations}
\end{eqnarray}
%%%%%%%%%%%%%%%%%%%%%%%%%%%%%%%%%%%%%%%%%%%%%%%%%
Note that, for arbitrary $D$, there are $D-2$ transverse polarizations $\lambda$.
For $D=4$, one can take
%%%%%%%%%%%%%%%%%%%%%%%%%%%%%%%%%%%%%%%%%%%%%%%%%
\begin{eqnarray}
\varepsilon_{\lambda} &=& \frac{1}{\sqrt{2}}\;
\left( \begin{array}{c}
  1 \\
  i \lambda
  \end{array}
\right),
\end{eqnarray}
%%%%%%%%%%%%%%%%%%%%%%%%%%%%%%%%%%%%%%%%%%%%%%%%%
with $\lambda=\pm 1$, so that $\lambda$ coincides with the light-front helicity of the gluon.

%%%%%%%%%%%%%%%%%%%%%%%%%%%%%%%%%%%%%%%%%%%%%%%%%%%%%%%%%%%%%%%%%%%%%%%%%%%%%%%%%%%%%%%%%%%%%%%%%%%%%%%%%%%%%%%%%%%%%%
%%%%%%%%%%%%%%%%%%%%%%%%%%%%%%%%%%%%%%%%%%%%%%%%%%%%%%%%%%%%%%%%%%%%%%%%%%%%%%%%%%%%%%%%%%%%%%%%%%%%%%%%%%%%%%%%%%%%%%
%%%%%%%%%%%%%%%%%%%%%%%%%%%%%%%%%%%%%%%%%%%%%%%%%%%%%%%%%%%%%%%%%%%%%%%%%%%%%%%%%%%%%%%%%%%%%%%%%%%%%%%%%%%%%%%%%%%%%%
%%%%%%%%%%%%%%%%%%%%%%%%%%%%%%%%%%%%%%%%%%%%%%%%%%%%%%%%%%%%%%%%%%%%%%%%%%%%%%%%%%%%%%%%%%%%%%%%%%%%%%%%%%%%%%%%%%%%%%

\end{document}